\documentclass[10pt, letter]{IEEEtran}

\usepackage[T1]{fontenc}
\usepackage{amssymb}
\usepackage{graphicx}
\usepackage{booktabs}
\usepackage{epstopdf}
\usepackage[cmex10]{amsmath}
\usepackage{makecell}
\usepackage{amsthm}
\usepackage{latexsym,bm}
\usepackage{color}
\usepackage{subfigure}
\usepackage{multirow}
\usepackage{longtable}
\usepackage{mathrsfs}
\usepackage{hyperref}
\usepackage{xcolor}
\hypersetup{colorlinks=true,
	linkcolor=black,
	anchorcolor=red,
	citecolor=blue,
	filecolor=red,
	urlcolor=red,
	pdfauthor=author}
\usepackage{tabularx}
\usepackage{cite}
\usepackage{threeparttable}
\usepackage{verbatim}
\makeatletter
\newif\if@restonecol
\makeatother

	\usepackage{algorithm}%
	\usepackage{algpseudocode}%
	\usepackage{amsmath}

	\theoremstyle{plain}

	\theoremstyle{definition}
	\theoremstyle{definition}

	\graphicspath{{figure/}}

\begin{document}
		\title{Low-Complexity Successive-Cancellation List Decoding of $2\times2$ Kernel Non-Binary Polar Codes}

		\author{Xinyu Zhou,   
Pingping Chen 
\thanks{X. Zhou and  P. Chen are with the College of Physics and Information Engineering, Fuzhou University, Fuzhou 350108, China.}}
		\maketitle
		
		\begin{abstract}

			Non-binary successive cancellation list (NB-SCL) decoding expands each surviving path into $q$ candidate branches at every information symbol, which causes high path expansion, sorting, and pruning complexity.  To address this issue, this paper proposes low-complexity list decoding algorithms for $2\times2$ kernel non-binary polar codes (NBPCs). First, we design a split-reduced non-binary successive cancellation list (SR-NBSCL) decoder that skips path splitting when the current symbol is sufficiently reliable. 
			We then exploit the final Rate-1 node structure and switch the last group of information symbols to simplified non-binary successive cancellation (NB-SC) decoding, resulting in the enhanced split-reduced non-binary successive cancellation list (ESR-NBSCL) decoder. 
			To further reduce branch expansion at unreliable symbols, we introduce an accumulated reliability-deviation (ARD) metric and propose an adaptive branch-pruning non-binary successive cancellation list (ABP-NBSCL) decoder, which prunes unreliable candidate branches before sorting and then  reducs the dominant sorting complexity. 
			Simulation results show that the proposed decoders achieve frame-error-rate (FER) performance close to that of conventional NB-SCL decoding with much lower complexity. 
			In particular, the ABP-NBSCL decoder reduces the path splitting number (PSN) by more than $80\%$ at several tested signal-to-noise ratios (SNRs), with  a negligible  performance loss.  
		\end{abstract}
		
		
		\vspace{-0.2cm}
		
		\section{Introduction}
		
		Polar codes, first discovered by Arikan, are the first capacity-achieving codes for binary-input discrete memoryless channels with an explicit and deterministic structure \cite{Arikan2009,Zhang2016}. Polar codes have also been applied to finite-length packet recovery in unsourced random access, multi-user spatial modulation systems, and probabilistically-shaped polar-coded modulation for 6G systems \cite{Shi2026EnhancedPilot,Xie2025PolarizationAided,Wu2025JointDesignPSPCM,Gao2026LearningDecodeDPC,Cui2025PolarCodedRSMA}. Under successive cancellation (SC) decoding, polar codes asymptotically achieve capacity with complexity $O(N\log N)$ for  the block length $N$ \cite{Arikan2009}. For finite-length codes, successive cancellation list (SCL) decoding outperforms SC decoding  and  approaches maximum-likelihood (ML) decoding at high signal-to-noise ratios (SNRs) \cite{Liu2026ParityConsistent}, but with the increased complexity   $O(LN\log N)$  and $L$ is the list size \cite{TalVardy2015,Zhang2016}. Moreover, Cyclic-redundancy-check (CRC)-aided SCL decoding can further  improve the   error rate  performance \cite{NiuChen2012,Yuan2025SoftOutputSCL}. 
		
		However, the increased decoding complexity for SCL with larger values of $L$ makes polar codes less attractive for practical purposes \cite{Xie2024PolarizationAidedNOMA}.  
		This   issue becomes more pronounced for non-binary polar codes (NBPCs) over $\mathbb{G}\mathbb{F}(q)$, since each non-binary information symbol has $q$ possible decisions \cite{Sasoglu2009,MoriTanaka2014}. 
		Under similar decoding algorithm and coding rate, non-binary NBPCs often performs   superior to their binary counterparts    \cite{ParkBarg2013,GoldinBurshtein2018,Chiu2022}. 
		Recent studies on NBPCs   focus on code construction, kernel design, and efficient SC-based decoding \cite{Karakchieva2020,Presman2016,ChenBaiMa2023,Qiu2024PolarCodedGMACPNC}.

Low-complexity SCL decoding has been extensively studied for binary polar codes.   Zhang et al. proposed a split-reduced SCL decoder that skips splitting when the current unfrozen bit is sufficiently reliable \cite{Zhang2016}. Other works further introduced reliability tests, path-decision rules, tree pruning, partitioned list decoding, and fast SC/SCL decoding to reduce redundant path operations \cite{LiDuChen2019,Gao2019,Wang2021,Chen2016TreePruning,Yuan2024SCOS,YaoMa2025}. Quantized-soft-information GRAND decoding has also been investigated for CRC-concatenated polar codes, showing lower time and memory complexity than CA-SCL decoding under short-code settings \cite{Yuan2023DSGRAND}. For short CRC-polar codes, a guessing-flipping framework further combines GRAND-generated error patterns with an SCL baseline decoder, where unreliable bits are flipped before SCL decoding to improve short-code performance under moderate complexity \cite{Li2026GuessingFlipping}.
However, reducing the complexity of NB-SCL decoding for NBPCs remains more challenging. Different from binary SCL decoding, where each information bit produces only two candidate branches, an information symbol over $\mathrm{GF}(q)$ expands each surviving path into $q$ candidate branches. Then, the number of path-metric candidates increases with both the list size $L$ and the field order $q$, and the dominant decoding complexity is on the order of $O(LqN\log N)$. Although a larger field order can improve the symbol representation capability of NBPCs, it also enlarges the candidate set generated during list decoding \cite{Xiong2016SymbolDecision}. Therefore, for low-complexity NB-SCL decoding, the key bottleneck lies in how to reduce unnecessary symbol-level branch expansion before path-metric sorting and pruning.

For NBPCs, most existing low-complexity decoding methods focus on tree- or node-level simplification. Yuan and Steiner extended  SC and SCL decoding to polar codes constructed from $2\times2$ non-binary kernels, and proposed a pruned-tree SCL decoder to reduce the  decoding complexity \cite{YuanSteiner2018}. Feng et al. simplified NB-SCL decoding by exploiting Rate-1 nodes, where only unreliable symbols inside a Rate-1 node are selected for candidate-path generation \cite{Feng2020}. Farsiabi et al. further developed a fast SC decoder for $2\times2$ kernel NBPCs by identifying structured constituent nodes with specific frozen/information-symbol patterns, avoiding the full traversal of the decoding tree \cite{Farsiabi2024}. These methods effectively reduce tree traversal or simplify specific constituent nodes. However, they do not fully address the symbol-level $q$-ary branch expansion and the resulting path-metric sorting burden in general NB-SCL decoding.

Motivated by this limitation, this paper proposes a family of low-complexity SCL decoding algorithms for $2\times2$ kernel NBPCs. The key idea is to reduce unnecessary $q$-ary branch expansion at the symbol level before path-metric sorting. For reliable information symbols, the decoder directly makes hard decisions without expanding all $q$ candidates. For unreliable symbols, candidate branches are generated, while highly unreliable branches are pruned before sorting. In addition, the final Rate-1 part of the decoding tree is processed by simplified NB-SC decoding to avoid repeated list expansion.

		The main contributions of this paper are summarized as follows.
		
		1) We propose a split-reduced non-binary SCL (SR-NBSCL) decoder based on the observation that path splitting is unnecessary for highly reliable non-binary information symbols. Gaussian-approximation (GA)-based thresholds are used to determine whether each surviving path should split or directly make a hard decision. Furthermore, by exploiting the final Rate-1 node structure, we develop an enhanced SR-NBSCL (ESR-NBSCL) decoder. The split-reduced rule is applied before $u_{N-K_1+1}$, while the remaining symbols $(u_{N-K_1+1},\ldots,u_N)$ are decoded by simplified NB-SC without list expansion.
		
		2) To further reduce the sorting complexity caused by   non-binary branch expansion, an adaptive branch-pruning non-binary SCL  (ABP-NBSCL) decoder is proposed. It introduces an accumulated reliability deviation metric to evaluate candidate branches and prune unreliable branches before path-metric sorting, thereby reducing the number of branches involved in sorting and pruning.

		3) We provide a complexity analysis for the proposed decoders in terms of the effective list size and the effective branch factor. The result shows that SR-NBSCL, ESR-NBSCL, and ABP-NBSCL can reduce the dominant $O(LqN\log N)$ complexity of conventional NB-SCL decoding. Simulation results  show that the ABP-NBSCL reduces the path splitting number (PSN) by more than $80\%$ under the tested settings with almost the same performance as the NB-SCL.
		
		The rest of this paper is organized as follows. Section~\ref{preliminaries} introduces the basic construction and decoding of NBPCs. Section~\ref{proposed} presents the SR-NBSCL and ESR-NBSCL decoders. Section~\ref{ABP-NBSCL} describes the ABP-NBSCL decoder. Section~\ref{simulation} gives the simulation results and complexity analysis. Section~\ref{conclusion} concludes this paper.

		\section{Preliminaries}
		\label{preliminaries}

		\subsection{Encoding Scheme for Non-Binary Polar Codes}
		
		%
		%
		The NBPC$(N,K)$ denotes a non-binary polar code    with symbol-length $N=2^n$, message length $K$, and rate $R=K/N$   over the non-binary Galois Field $\mathbb{GF}(q)$ with $q=2^p$ and $p>1$. The elements of $\mathbb{GF}(q)$ are represented as $\alpha^{-\infty}=0, \alpha^0, \alpha^1, \dots, \alpha^{q-2}$, where $\alpha$ is the root of a primitive polynomial $f(x) = \alpha_0 + \alpha_1 x + \alpha_2 x^2 + \dots + \alpha_p x^p$, $\alpha_i \in \mathbb{GF}(2)$. Each element in $\mathbb{GF}(q)$ corresponds to a binary vector of length $p$. An information bit sequence $\bm{b} = [b_0, \dots, b_{K_b-1}]^T$ of length $K_b = pK$ can be converted into a non-binary message vector $\bm{m} =[m_0, \dots, m_{K-1}]^T$ of length $K$ over $\mathbb{GF}(q)$.
		
		In the NBPC encoder, the elements of $m$ are placed at $K$ information positions of an input sequence $\bm{u} =[u_0, \dots, u_{N-1}]^T$. The remaining positions are frozen and filled with $0$ symbols. We denote the sets of information and frozen positions as $\mathcal{A}$ and $\mathcal{A}^c$, respectively. Then, the vector $u$ is encoded into a non-binary polar codeword $\bm{x} = [x_0, \dots, x_{N-1}]^T$ using the following equation
		\begin{equation}
			x^T = u^T G_2^{\otimes n},
		\end{equation}
		where $\otimes n$ denotes the $n$-th Kronecker power and $G_2^{\otimes n}$ is the generator matrix of the NBPC. The kernel $G_2$   used in this paper is the extension of $2 \times 2$ Arikan kernel to the non-binary Galois field \cite{YuanSteiner2018}, defined as
		\begin{equation}
			\mathbf{G}_2 = \begin{bmatrix} \mu & \quad 0 \\[1ex] \gamma & \quad \delta \end{bmatrix}.
		\end{equation}
		where $\mu, \gamma, \delta \in \mathbb{G}\mathbb{F}(q) \setminus 0$, denoting kernel coefficients, are non-zero elements in $\mathbb{G}\mathbb{F}(q)$.

		\subsection{SC And SCL Decoding of NBPC}
		
\begin{figure}[!t]
			\centering
			\includegraphics[width=0.55\linewidth]{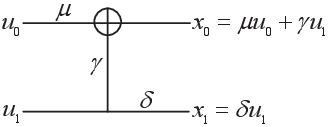}
			\vspace{-0.15cm}
			\caption{\label{fig:basic_unit} Non-binary $2 \times 2$ kernel with $\mu, \gamma, \delta \in \mathbb{G}\mathbb{F}(q) \setminus 0$.}
			\vspace{-0.5cm}
		\end{figure}

		Fig. \ref{fig:basic_unit} illustrates the basic polarization unit based on the $2 \times 2$ kernel, which encodes the symbols $(u_0, u_1)$ into $(x_0, x_1) = (\mu u_0 + \gamma u_1, \delta u_1)$. Both addition and multiplication operations are performed over $\mathbb{G}\mathbb{F}(q)$. Next, we show the decoding update rules of NBPCs based on the kernel $G_2$.

		For log-likelihood ratios (LLR)-based SC and SCL decoding, each codeword symbol has $q=2^p$ possible values. Unlike binary polar codes using a scalar bit-LLR, a non-binary symbol requires an LLR vector to represent the reliabilities of all possible $q=2^p$ field elements. During channel polarization, the transition probability of the $i$-th synthetic channel is denoted by $W_N^{(i)}(y_1^N,\hat{u}_1^{i-1}|u_i)$, where $y_1^N$ is the received channel-output sequence and $\hat{u}_1^{i-1}$ denotes the previously decoded symbols. For a candidate symbol value $\lambda\in\mathrm{GF}(q)$, the corresponding LLR is defined a
		 \begin{equation}
			 	L(u_i)_{\lambda} = \ln \frac{W_N^{(i)}(y_1^N, \hat{u}_1^{i-1} \mid u_i = 0)}{W_N^{(i)}(y_1^N, \hat{u}_1^{i-1} \mid u_i = \lambda)}, \quad \lambda \in \mathbb{G}\mathbb{F}(q).
			 	\label{eq:llr_init}
			 \end{equation}

		The symbol LLRs in \eqref{eq:llr_init} can either be calculated from received $p$-ary constellation symbols or extracted from bit  LLRs. For the latter case, let the channel bit LLRs corresponding to $N_b = pN$ different bits be denoted by $\mathcal{L}_k$, $0 \le k \le N_b - 1$. Let the binary representation of $\lambda\in\mathrm{GF}(q)$ be $(\lambda(0),\ldots,\lambda(p-1))$. Then, the initial symbol-LLR for the $i$-th codeword symbol and the corresponding bit LLRs  have the   relation \cite{Farsiabi2024}
		\begin{equation}
			L_{i}^{(init)} = \sum_{j=0}^{p-1} \Big( \lambda(j) \oplus \mathrm{HD}(\mathcal{L}_{ip+j}) \Big) |\mathcal{L}_{ip+j}|, \quad 0 \le i \le N-1,
			\label{eq:llr_calculation}
		\end{equation}
		where $\oplus$ is the XOR operator and $\mathrm{HD}(\cdot)$ makes hard decision on bit-LLRs as
		\begin{equation}
			\mathrm{HD}(\eta) = 
			\begin{cases} 
				0, & \text{if } \eta > 0; \\ 
				1, & \text{otherwise}. 
			\end{cases}
			\label{eq:hd_function}
		\end{equation}

\begin{figure}[!t]
			 	\centering
			 	\includegraphics[width=0.9\linewidth]{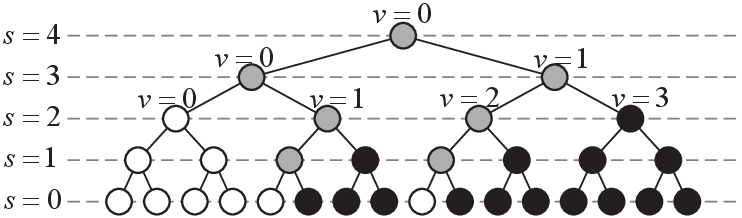}
			 	\vspace{-0.15cm}
			 	\caption{\label{fig:nbpc_tree} Binary tree for non-binary polar codes with $N=16$ and $K=8$. (Black nodes and white nodes represent information symbols and frozen symbols, respectively).}
			 	\vspace{-0.5cm}
			 \end{figure}

		Moreover, the decoding process of NBPCs can be represented as message passing over a binary decoding tree \cite{YuanSteiner2018}. Fig. \ref{fig:nbpc_tree} shows a binary decoding tree for an NBPC (16, 8), where $s$ denotes the level in the tree and $0 \le v \le 2^{n-s}-1$ is the node index from left to right. $N_s = 2^s$ is used to denote the length of a node rooted at level $s$. The nodes of the tree can be identified with $(v,s)$ pair where each one, except the leaf nodes, has two children of the left child-node $(2v,s-1)$ and the right child-node $(2v+1,s-1)$.

		Let the soft information associated with the $i$-th symbol of node $(v,s)$ be a length-$q$ vector $\bm{L_{i}}^{(v,s)} =[l_{i,0}^{(v,s)}, \dots, l_{i,q-1}^{(v,s)}]$, for $0 \le i \le N_s-1$. Node $(v,s)$ receives a $q \times N_s$ soft information matrix $\mathbf{L}^{(v,s)} =[\bm{L_0^{(v,s)}}, \dots, \bm{L_{N_s-1}^{(v,s)}}]$ from its parent node. It performs a symbol hard decision to obtain the estimated vector $\bm{\chi^{(v,s)}} =[\chi_0^{(v,s)}, \dots, \chi_{N_s-1}^{(v,s)}]$. The estimated vector $\chi^{(v,s)}$ is then passed to its parent node $(\lfloor v/2 \rfloor, s+1)$. Following these definitions, the root node $(0,n)$ receives the channel messages $\mathbf{L}^{(0,n)} =[\bm{L_0}, \dots, \bm{L_{N-1}}]$. It finally outputs the estimated NBPC codeword $\bm{\chi^{(0,n)}} =[\chi_0^{(0,n)}, \dots, \chi_{N-1}^{(0,n)}]$ at the leaf nodes, where $N=2^n$.

\begin{figure}[!t]
			\centering
			\subfigure[]{
				\label{fig:f_func_graph}
				\includegraphics[width=0.45\columnwidth]{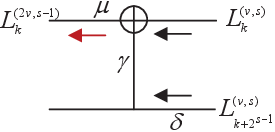}
			}
			\hfill
			\subfigure[]{
				\label{fig:g_func_graph}
				\includegraphics[width=0.45\columnwidth]{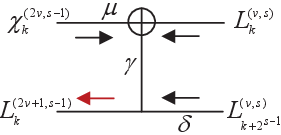}
			}
			\vspace{-0.15cm}
			\caption{\label{fig:fg_functions} Two types of messages calculated at node $(v,s)$: (a) message toward the left child (F-function), and (b) message toward the right child (G-function).}
			\vspace{-0.5cm}
		\end{figure}

Upon receiving $\mathbf{L}^{(v,s)}$, each non-leaf node updates the soft information of its two child nodes in a successive manner. 
The left-child message $\mathbf{L}^{(2v,s-1)}$ is first computed by the $F$-function. 
After the left-child estimate is obtained, the right-child message $\mathbf{L}^{(2v+1,s-1)}$ is computed by the $G$-function, as illustrated in Fig.~\ref{fig:fg_functions}. 
Let $\bm{L}_{k}^{(2v,s-1)}=[l_{k,0}^{(2v,s-1)},\ldots,l_{k,q-1}^{(2v,s-1)}]$ denote the LLR vector of the $k$-th symbol in the left child, where $0\leq k\leq 2^{s-1}-1$. 
Then, the $F$-function is expressed as
 
		\begin{equation}
			\begin{split}
				l_{k,\lambda}^{(2v,s-1)} &= \sum_{u_t \in \mathbb{G}\mathbb{F}(q)} \exp \Big( \big( l_{k,\gamma u_t}^{(v,s)} + l_{k+2^{s-1},\delta u_t}^{(v,s)} \big) \\
				&\qquad\qquad\qquad - \big( l_{k,\lambda+\gamma u_t}^{(v,s)} + l_{k+2^{s-1},\delta u_t}^{(v,s)} \big) \Big), \\
				&\qquad\qquad\qquad\qquad\qquad 0 \le k \le 2^{s-1}.
			\end{split}
			\label{eq:f_func_exact}
		\end{equation}
		We can use the following equation to simplify   \eqref{eq:f_func_exact}:
		\begin{equation}
			\ln \left( \sum_i e^{-f_i} \right) \approx -\min_i (f_i).
			\label{eq:ems_approx}
		\end{equation}
		We then  obtain a simplified approximation for \eqref{eq:f_func_exact}   as
		\begin{equation}
			\begin{split}
				l_{k,\lambda}^{(2v,s-1)} &\approx \min_{u_t \in \mathbb{G}\mathbb{F}(q)} \left( l_{k,\lambda+\gamma u_t}^{(v,s)} + l_{k+2^{s-1}, \delta u_t}^{(v,s)} \right) \\
				&\quad - \min_{u_t \in \mathbb{G}\mathbb{F}(q)} \left( l_{k,\gamma u_t}^{(v,s)} + l_{k+2^{s-1}, \delta u_t}^{(v,s)} \right), 
			\end{split}
			\label{eq:f_func}
		\end{equation}
		where $u_t = u_k^{(2v+1,s-1)}$.
		
		After receiving the estimated output from the left child, the node calculates $\mathbf{L}^{(2v+1,s-1)}$ using the G-function. Referring to Fig. \ref{fig:fg_functions}(b), the G-function is expressed as
		\begin{equation}
			\begin{aligned}
				l_{k+2^{s-1},\lambda}^{(2v+1,s-1)} \approx &\; l_{k,\chi_k^{(2v,s-1)}+\gamma \lambda}^{(v,s)} + l_{k+2^{s-1},\delta \lambda}^{(v,s)} \\
				&- l_{k,\chi_k^{(2v,s-1)}}^{(v,s)} - l_{k+2^{s-1},0}^{(v,s)}.
			\end{aligned}
			\label{eq:g_func_approx_new}
		\end{equation}
		where the addition and multiplication are implemented over $\mathbb{G}\mathbb{F}(q)$.
		
		Once all leaf nodes receive the soft information from their parents, the symbol $u_i$ at the $i$-th leaf node is estimated as

		\begin{equation}
			\hat{u}_i = 
			\begin{cases} 
				0, & \text{if } i \in \mathcal{A}^c; \\ 
				\mathop{\arg\min}\limits_{\lambda \in \mathbb{G}\mathbb{F}(q)} (l_{i,\lambda}), & \text{otherwise}. 
			\end{cases}
			\label{eq:hard_decision}
		\end{equation}

For the $j$-th decoding path, where $j\in\{1,2,\ldots,L\}$, the path metric at the $i$-th information symbol is updated as

		Then, we can decode the information symbols from \eqref{eq:hard_decision}  by  a hard decision. It selects  the field element with the minimum LLR value in each LLR vector, which corresponds to the maximum-likelihood symbol decision.
		
Unlike NB-SC decoding, which makes a single hard decision for each symbol based on the minimum LLR value, NB-SCL decoding considers all $q$ possible values of each information symbol $u_i$ for every surviving path. 
  At each decoding stage, only the $L$ paths with the minimum path metric ($\mathrm{PM}$) values survive  to reduce the complexity.
		
		For the $j$-th decoding path, $j \in \{1, 2, \dots, L\}$, at the $i$-th information symbol, we update the $\mathrm{PM}$  as
		\begin{equation}
			\mathrm{PM}_j^{(i)} = 
			\begin{cases} 
				\mathrm{PM}_j^{(i-1)},  \ \text{if } \hat{u}_i[j] = \arg\min(L_i), \\ 
				\mathrm{PM}_j^{(i-1)} + \left| l_{i,\eta} - \min\limits_{\lambda \in \mathbb{G}\mathbb{F}(q) \setminus \eta} l_{i,\lambda} \right|,\  \text{otherwise}.
			\end{cases}
			\label{eq:pm_update}
		\end{equation}
		where $\eta \in \mathbb{G}\mathbb{F}(q)$ denotes the estimated symbol of the current path at the $i$-th information position.
		\section{Proposed Split-Reduced SCL Decoding Algorithm for NBPCs}
		\label{proposed}

This section presents a reliability-based path-splitting rule for non-binary symbols over $\mathrm{GF}(q)$. Based on this rule, a split-reduced NB-SCL decoder is developed to avoid unnecessary path expansion at highly reliable information symbols and reduce the decoding complexity of conventional NB-SCL decoding.

		\subsection{Path Splitting Rule}
		
		By channel polarization, each information symbol $u_i\in\mathbb{G}\mathbb{F}(q)$ is transmitted through the synthetic sub-channel $W_N^{(i)}(y_1^N,u_1^{i-1}|u_i)$. The decoding reliability of $u_i$ is determined by this synthetic sub-channel. The Bhattacharyya parameter provides a tractable reliability measure for each synthetic channel  and  is recursively updated during polar code construction.  However, this  scalar reliability measure is not directly applicable to general $q$-ary channels. Following the split-reduced decoding principle in \cite{Arikan2009}, we use symbol-wise  \textit{a posteriori} probability to define the reliability metric. 
		The symbol error probability of each synthetic sub-channel can be estimated by Gaussian approximation (GA) or density evolution. 
		
		Consider the synthetic sub-channel $W_N^{(i)}$ and an input symbol $u_i \in \mathbb{G}\mathbb{F}(q)$. 
		Assume that all previous symbols are decoded correctly. 
		Let $P_e(u_i)$ denote the estimated error probability for deciding $u_i$. 
		The probability is averaged over all possible pairs $(y_1^N,u_1^{i-1})$. 
		Then, $P_e(u_i)$ is given by 
		\begin{equation}
			\begin{split}
				P_e(u_i) &= P(\hat{u}_i \neq u_i) \\
				&= \sum_{u_1^{i-1} \in \mathcal{X}^{i-1}} \sum_{y_1^N \in \mathcal{Y}^N} P \Big( \hat{u}_i(y_1^N, u_1^{i-1}) \neq u_i \\
				&\qquad\qquad \mid \hat{u}_1^{i-1} = u_1^{i-1}, u_i, y_1^N \Big).
			\end{split}
			\label{eq:error_prob}
		\end{equation}
		where $\mathcal{X}$ represents the symbol set over $\mathbb{G}\mathbb{F}(q)$ and $\mathcal{Y}$ denotes the received signal set. Thus, $P_e(u_i)$ measures the probability that the SC decision on $u_i$ is incorrect when the previous symbols are known correctly.
		
		With $P_e(u_i)$, we can have $1-P_e(u_i)$ as the confidence level for the $i$-th sub-channel. Direct evaluation of \eqref{eq:error_prob} is generally difficult, so $P_e(u_i)$ is computed offline by Monte-Carlo simulation. 
		Following \cite{TalVardy2013}, we formulate the LLR recursion based on the non-binary $2 \times 2$ kernel as
		\begin{equation}
			\begin{split}
				L_N^{(2i-1)}(y_1^N, u_1^{2i-2}) &= L_{N/2}^{(i)} (y_1^{N/2}, \mu u_{1,o}^{2i-2} \oplus \gamma u_{1,e}^{2i-2}) \\
				&\quad \boxplus_{\mathbb{G}\mathbb{F}(q)} L_{N/2}^{(i)} (y_{N/2+1}^N, \delta u_{1,e}^{2i-2}),
			\end{split}
			\label{eq:llr_recursion_odd}
		\end{equation}
		and
		\begin{equation}
			\begin{split}
				L_N^{(2i)}(y_1^N, u_1^{2i-1}) &= L_{N/2}^{(i)} (y_1^{N/2}, \mu u_{1,o}^{2i-2} \oplus \gamma u_{1,e}^{2i-2}) \\
				&\quad + L_{N/2}^{(i)} (y_{N/2+1}^N, \delta u_{1,e}^{2i-2}).
			\end{split}
			\label{eq:llr_recursion_even}
		\end{equation}
	where $a \boxplus b
	= \log \frac{1+e^{a+b}}{e^a+e^b}$, and the operation is performed over $\mathbb{G}\mathbb{F}(q)$. Then, we can have the expectation of  \eqref{eq:llr_recursion_odd} and \eqref{eq:llr_recursion_even}  as
		\begin{equation}
			\mathbb{E}[\mathbf{L}(u_1)] = \Phi_q^{-1} \Big( \Phi_q(\mathbb{E}[\mathbf{L}(y_1)]) \cdot \Phi_q(\mathbb{E}[\mathbf{L}(y_2)]) \Big),
		\end{equation}
		\begin{equation}
			\mathbb{E}[\mathbf{L}(u_2)] = \mathbb{E}[\mathbf{L}(y_1)] + \mathbb{E}[\mathbf{L}(y_2)].
		\end{equation}
		where $\mathbb{E}$ denotes expectation and
		\begin{equation}
			\Phi_q(x) \!=\! 
			\begin{cases} 
				\begin{aligned}
					&1 - \textstyle\idotsint \bigl[ \log_q \bigl( 1 + \sum_{k=1}^{q-1} e^{-u_k} \bigr) \bigr] \\
					&\quad \times \prod_{k=1}^{q-1} \tfrac{1}{\sqrt{4\pi x}} e^{-\frac{(u_k-x)^2}{4x}} \, du_1 \cdots du_{q-1},
				\end{aligned} & x>0,\\[1ex]
				0, & x=0.
			\end{cases}
			\label{eq:phi_q}
		\end{equation}

		The $(q-1)$-fold integral in \eqref{eq:phi_q} is expensive to evaluate and can be approximated   by Monte-Carlo integration as
		\begin{equation}
			\Phi_q(x) \approx 
			\begin{cases} 
				1 - \frac{1}{M} \sum\limits_{j=1}^{M} \log_q \left( 1 + \sum\limits_{k=1}^{q-1} e^{-u_{k,j}} \right), & x > 0, \\ 
				0, & x = 0. 
			\end{cases}
			\label{eq:phi_q_approx}
		\end{equation}
		where $M$ is the number of Monte-Carlo samples, and $u_{k,j}$ denotes the $k$-th component of the $j$-th sampled LLR vector.
		
		For a $q$-ary symbol, an error occurs if the correct symbol is confused with one of the other $q-1$ symbols. 
		We approximate the symbol error probability by the union bound of these pairwise error events. 
		Let $m_i^{\rm GA}$ denote the mean of the pairwise LLR  obtained from the GA for $W_N^{(i)}$. 
		Then, \eqref{eq:error_prob} can be rewritten as
		\begin{equation}
			P_e(u_i) \approx (q-1)Q\left(\sqrt{\frac{m_i^{\rm GA}}{2}}\right),
		\end{equation}
		where $Q(x)=\frac{1}{\sqrt{2\pi}}\int_x^{+\infty}e^{-t^2/2}dt$. 
		The factor $(q-1)$ accounts for the $q-1$ competing symbols other than the correct one. 
		For a fixed noise variance $\sigma_n^2$, these values are computed offline and used for code construction and the frozen-symbol selection.
		
		Having defined the reliability metric and threshold, we propose the following path splitting rule:
		\begin{equation}
			\begin{aligned}
				P_l(u_i=0 \mid y_1^N, \hat{u}_1^{i-1}) &> 1 - P_e(u_i) \\
				P_l(u_i=1 \mid y_1^N, \hat{u}_1^{i-1}) &> 1 - P_e(u_i) \\
				&\vdots \\
				P_l(u_i=q-1 \mid y_1^N, \hat{u}_1^{i-1}) &> 1 - P_e(u_i).
			\end{aligned}
			\label{eq:splitting_rule_prob}
		\end{equation}

		If any of these inequalities holds, the $l$-th decoding path survives without splitting. Otherwise, it splits into $q$ candidate branches. We apply Bayes' theorem to simplify this splitting rule in the  LLR domain:
		\begin{equation}
			\hat{u}_i = 
			\begin{cases} 
				0, & \text{if } \min(LLR_{0,1}, \dots, LLR_{0,q-1}) > T, \\ 
				1, & \text{if } \min(LLR_{1,0}, \dots, LLR_{1,q-1}) > T, \\ 
				\vdots \\ 
				q-1, & \text{if } \min(LLR_{q-1,0}, \dots, LLR_{q-1,q-2}) > T.
			\end{cases}
			\label{eq:splitting_rule_llr}
		\end{equation}
		where $T$ is the decision threshold given by $T = \log \left( \frac{1-P_e(u_i)}{P_e(u_i)} \right)$. The term $LLR_{k,j}$ is defined as $LLR_{k,j} = \log \left( \frac{P_l(u_i=k \mid y_1^N, \hat{u}_1^{i-1})}{P_l(u_i=j \mid y_1^N, \hat{u}_1^{i-1})} \right)$ , $j \neq k$.
		%
		%
		\subsection{The Decoding Algorithm}
		
		In NB-SCL decoding, each surviving path is extended to $q$ candidate paths at each information symbol. 
		This  causes high sorting and pruning complexity. The reliability measure defined in \cite{Zhang2016} gives a way to decide whether path splitting is necessary. 
		Based on this measure, we   state two conjectures on the behavior of correct and incorrect paths.
		
		\textbf{Conjecture 1:} 
		Suppose that the correct decoding path survives until $u_1^{i-1}$. 
		Under the GA, as $P_e(u_i)$ approaches zero, with high probability, the current path will survive at $u_i$ without splitting and $u_i$ will be correctly decoded. 
		In addition, with the increasing reliability of the subsequent subchannels corresponding to $u_{i+1}^{N}$, the correct path will survive till termination without splitting with high probability.
		
		\textit{Proof:} See Appendix~A.
		
		\textbf{Conjecture 2:} Consider any incorrect path surviving at symbol $u_i$, where at least one decision error occurred previously. This path splits with extremely high probability in the subsequent decoding stages $\{i+1, \dots, N\}$.
		
		\textit{Proof:} See Appendix B.
		
		According to these two conjectures, the correct path can pass reliable symbols without splitting with high probability. 
		An incorrect path is likely to split again at a subsequent reliable symbol. 
		Thus, the number of consecutive non-splitting decisions can be used to distinguish these two types of paths.
		
		We use a counter $w_l[i]$ for the $l$-th path at stage $i$. 
		If path $l$ satisfies (21) and does not split at stage $i$, then$w_l[i]=w_l[i-1]+1$
		If path $l$ splits, the counters of all its child paths are set to zero.

 Let $\omega$ denote a predefined counter threshold. If $w_l[i]>\omega$, path $l$ is considered reliable. 
When pruning is needed, the decoder  keeps the candidate paths satisfying $w_l[i]>\omega$, and the other paths are deleted.  
If more than $L$ paths satisfy this condition, only the $L$ paths with the smallest path metrics are kept. 
If no path satisfies this condition, the decoder falls back to the conventional pruning rule and retains the $L$ paths with the smallest path metrics.

		With this rationale, we propose the  split-reduced NB-SCL (SR-NBSCL) in Algorithm 1.
		\begin{algorithm}[ht]
			\caption{Split-Reduced NB-SCL Decoder}
			\label{alg:sr_nbscl}
			\begin{algorithmic}[1]
				\renewcommand{\algorithmicforall}{\textbf{for all}}
				
				\State \textbf{Step 1:} Initialize the decoder at $u_1$. Set the list size $L$, the counter threshold $\omega$, the path counters, and the initial path metrics.
				
				\State \textbf{Step 2:} For the $l$-th path at an information symbol $u_i$, evaluate \eqref{eq:splitting_rule_llr}. If it holds, make a direct decision for $\hat{u}_i$ without path splitting and increase $w_l[i]$ by one. Otherwise, split the path into $q$ candidate branches and reset the counters of the resulting branches to zero.
				
				\State \textbf{Step 3:} If the number of surviving paths exceeds $L$, perform pruning. Retain the paths with $w_l[i]>\omega$ if such paths exist; otherwise, retain the best $L$ paths according to the path metric (PM).
				
				\State \textbf{Step 4:} If $i<N$, set $i=i+1$ and return to Step 2. Otherwise, output the candidate codeword with the minimum PM value.
				
			\end{algorithmic}
		\end{algorithm}
		\subsection{The Enhanced Split-Reduced SCL Decoding Algorithm}

To further reduce the decoding complexity, we propose an enhanced split-reduced NB-SCL (ESR-NBSCL) decoder. We terminate list expansion once the decoder reaches a terminal Rate-1 information-symbol block, and then decode the remaining symbols using simplified NB-SC decoding. When the preceding information symbols are correctly decoded, this simplified decoding produces the same performance as maximum-likelihood (ML) decoding.

		The key   is to determine this starting Rate-1  node index. 
		For NBPCs constructed from $2\times2$ kernels, the SC decoding process can be described by a binary decoding tree. 
		Each leaf node corresponds to one symbol $u_i$. 
		A simplified NB-SC operation can be applied to a complete subtree only when all leaf nodes in this subtree are information symbols. 
		Therefore, we need to find a consecutive information-symbol block at the end of the decoding tree.
		
		We count the information symbols backward from the last symbol $u_N$. 
		The counting stops when the first frozen symbol is met. 
		Let the number of consecutive information symbols be $C$. 
		Since a complete subtree generated by a $2\times2$ kernel has length $2^k$, the switching length is also be a power of two. 
		Thus, we choose $K_1$ as the largest power of two not larger than $C$, i.e.,
		$K_1=2^k\le C$. 
		Then, the first symbol of this block is $u_{N-K_1+1}$. 
		The decoder uses the splitting rule before $u_{N-K_1+1}$ and applies simplified NB-SC decoding from $u_{N-K_1+1}$ to $u_N$.
		
		Consider a simple NBPC with length $N=8$ and rate $R=0.5$. 
		Suppose that the information set is $\{u_4,u_6,u_7,u_8\}$. 
		Counting backward from $u_8$ gives two consecutive information symbols, $u_7$ and $u_8$. 
		Thus, $C=2$ and $K_1=2$. 
		The starting index is $N-K_1+1=7$. 
		In this case, path splitting is avoided from $u_7$ onward.
		
		\textbf{Theorem 1.} Suppose the desired $K_1$ is determined and a genie correctly supplies all information symbols with the indices $\{i : 1 \le i \le N - K_1\}$. The NB-SC decoder achieves exactly the same performance as the ML decoder.
		
		\textit{Proof:} See Appendix C.
		
		For NBPCs constructed from $2\times2$ kernels, the last $K_1$ consecutive information symbols form a complete terminal node in the decoding tree. 
		This node contains no frozen symbol. 
		Thus, after the previous information symbols $u_1^{N-K_1}$ are correctly decoded, the remaining vector $(u_{N-K_1+1},\ldots,u_N)$ can be regarded as a length-$K_1$ non-binary subcode. 
		Since the kernel matrix $G_2^{\otimes k_1}$ is nonsingular over $\mathrm{GF}(q)$, the simplified NB-SC decoder can select the same suffix estimate as the ML decoder for this subcode. 
		This observation supports the use of simplified NB-SC decoding from $u_{N-K_1+1}$ without path splitting. Based on the above observation, the ESR-NB-SCL  decoding algorithm is proposed, given in Algorithm~\ref{alg:esr_nbscl}.  
		
		\begin{algorithm}[ht]
			\caption{Enhanced Split-Reduced NB-SCL Decoder}
			\label{alg:esr_nbscl}
			\begin{algorithmic}[1]
				
				\State \textbf{Step 1:} Initialize the decoder starting from the first symbol $u_1$.
				
				\State \textbf{Step 2:} For the $l$-th path at an information symbol $u_i$, if \eqref{eq:splitting_rule_llr} holds, make a direct hard decision for $\hat{u}_i$ without path splitting. Otherwise, split the decoding path into $q$ candidate branches. Update the counter $w_l[i]$ for each path simultaneously.
				
				\State \textbf{Step 3:} When the total number of paths exceeds the list size $L$, prune those paths with counters smaller than the threshold $\omega$. If no path exceeds $\omega$, select the best $L$ paths based on the path metric (PM).
				
				\State \textbf{Step 4:} If $i < N - K_1$, increment $i$ to $i + 1$ and return to Step 2. Otherwise, apply simplified NB-SC decoding to obtain a unique estimate sequence $(\hat{u}_{N-K_1+1}, \ldots, \hat{u}_N)$ for each surviving path. Output the candidate codeword with the minimum PM value.
				
			\end{algorithmic}
		\end{algorithm}
		
		After $K_1$ is determined, the enhanced decoder works in two parts. 
		For the information symbols before $u_{N-K_1+1}$, the proposed splitting rule in \eqref{eq:splitting_rule_llr} is used. 
		For the remaining information symbols $(u_{N-K_1+1},\ldots,u_N)$, a  NB-SC decoding is applied. 
		Thus, the path splitting is removed in the last $K_1$ information-symbol positions.
		
		We see that the decoding complexity is further reduced, since the enhanced decoder avoids list expansion and pruning after $u_{N-K_1+1}$. 
		With this simplification, the following theorem shows that the enhanced decoder does not degrade the performance of the original split-reduced NB-SCL decoder.
		
		\textbf{Theorem 2:} 
		The decoding error performance of the enhanced split-reduced NB-SCL decoder is no worse than that of the original split-reduced NB-SCL decoder.
		
		\textit{Proof:} See Appendix D.
		
		It is also useful to consider the worst case. 
		The worst case occurs when no path satisfies the LLR threshold or the counter threshold before $u_{N-K_1+1}$. 
		In this case, the decoder behaves like the original NB-SCL decoder before $u_{N-K_1+1}$. 
		However, the index $N-K_1+1$ always exists once the last consecutive information-symbol block is determined. 
		Thus, a NB-SC decoding can still be used for the last $K_1$ symbols. 
		Therefore, the enhanced decoder can reduce the complexity even in the worst case, without degrading the error performance.
		
		\section{Adaptive Branch Pruning Scheme Based on Accumulated Reliability Deviation}
		\label{ABP-NBSCL}
		
		This section presents an adaptive branch-pruning NB-SCL (ABP-NBSCL) decoder. The proposed decoder first identifies unreliable symbols from non-binary Rate-1 nodes and then employs an accumulated reliability-deviation (ARD) metric to prune unreliable candidate branches before path-metric sorting.
		
		\subsection{Construction of Candidate Set}
		
		Following the critical-set construction, the NBPCs is divided into multiple Rate-1 sub-codes, and the first information symbol of each sub-block is collected into the candidate set. Let $\mathcal{T}_{\mathrm{R1}}$ denote the set of these sub-codes.  
		For a node $B\in\mathcal{T}_{\mathrm{R1}}$, let
		\begin{equation}
			\mathcal{I}(B)=\{i_B,i_B+1,\ldots,i_B+M_B-1\}\subseteq\mathcal{A},
			\qquad M_B=2^{m_B}.
			\label{eq:r1_index_set}
		\end{equation}
		The first symbol in this node is denoted by $\kappa(B)=i_B$. 
		The candidate set (CS) is defined as
		\begin{equation}
			\mathrm{CS}=\{\kappa(B):B\in\mathcal{T}_{\mathrm{R1}}\}.
			\label{eq:cs_definition}
		\end{equation}
		The remaining information symbols are collected in $\mathcal{A}^r=\mathcal{A}\setminus\mathrm{CS}$, and the search set is
		\begin{equation}
			\mathcal{A}^u=\mathrm{CS}.
			\label{eq:au_definition}
		\end{equation}

		The reason for using \eqref{eq:cs_definition} is that the first symbol of a Rate-1 node carries the dominant error event of this node. 
		The following theorem gives  this property.
		
		\textbf{Theorem 3.} 
		Consider a non-binary Rate-1 node of length $M=2^m$ over $\mathbb{G}\mathbb{F}(q)$. 
		Let $\mathcal{E}_{\mathrm{R1}}$ be the event that this node is decoded incorrectly. 
		Let $\mathcal{E}_{0}$ be the event that the first information symbol of this node is decoded incorrectly. 
		If the local codeword-symbol errors are independent and each has probability $p<\epsilon$, then
		\begin{equation}
			\begin{aligned}
				0
				&\le
				P(\mathcal{E}_{\mathrm{R1}})-P(\mathcal{E}_{0})\\
				&\le
				\sum_{r=2}^{M}\binom{M}{r}p^r(1-p)^{M-r}\\
				&<
				\sum_{r=2}^{M}\binom{M}{r}\epsilon^r.
			\end{aligned}
			\label{eq:theorem3_bound}
		\end{equation}
		Thus, when the Rate-1 node is reliable, $P(\mathcal{E}_{0})$ is close to $P(\mathcal{E}_{\mathrm{R1}})$.
		
		\textit{Proof:} See Appendix E.
		
\begin{figure}[!t]
			 	\begin{minipage}[b]{0.48\textwidth}
				 		\centering
				 		\includegraphics[width=0.8\textwidth]{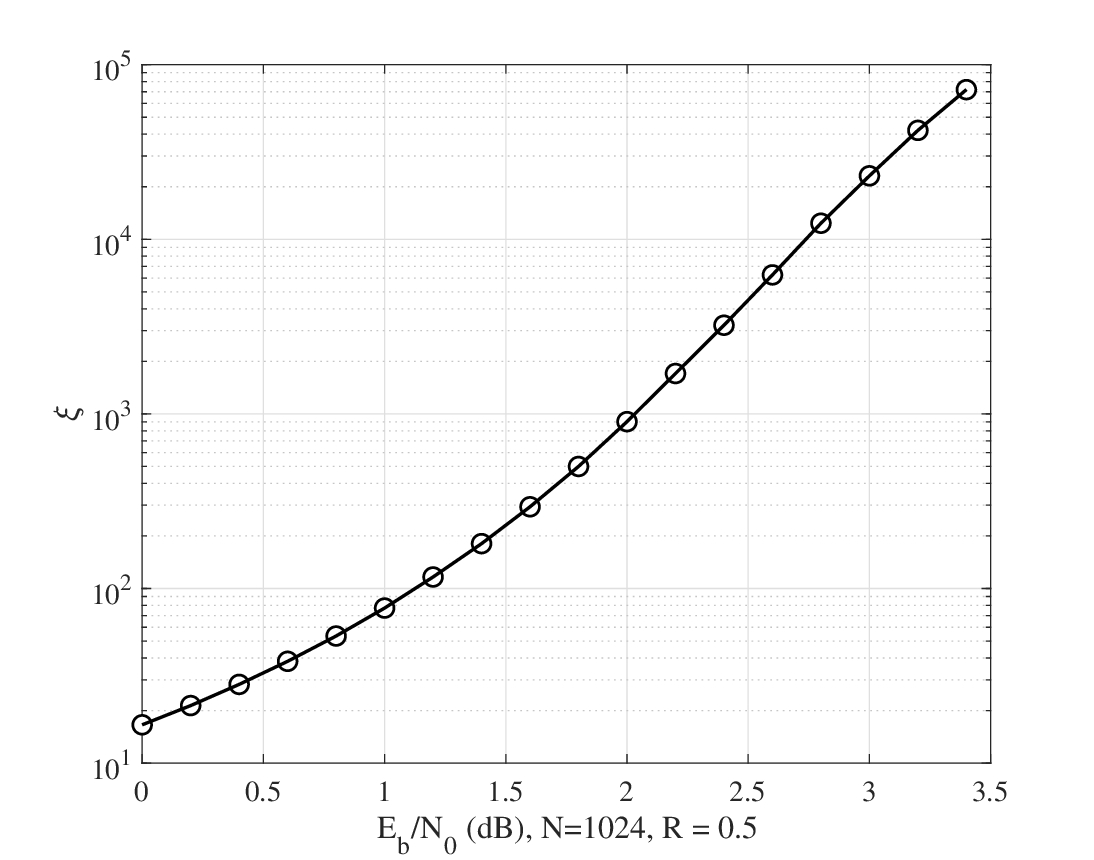}
				 		\vspace{-0.15cm}
				 		\caption{\label{fig:xi_SNR} Relationship between $\xi$ and SNR $E_b/N_0$.}
				 	\end{minipage}
			 	\vspace{-0.5cm}
		\end{figure}

		This result is consistent with the critical-set observation in \cite{Zhang2017Progressive}, where the first error is included in the critical set with probability higher than $99\%$ in simulations. 
		For NBPCs, we use the symbol error probabilities obtained by GA or Monte-Carlo construction to measure the concentration effect. 
		Define
		\begin{equation}
			\xi=
			\frac{
				\frac{1}{|\mathrm{CS}|}\sum\limits_{i\in\mathrm{CS}}P_e(W_N^{(i)})
			}{
				\frac{1}{|\mathcal{A}\setminus\mathrm{CS}|}
				\sum\limits_{i\in\mathcal{A}\setminus\mathrm{CS}}P_e(W_N^{(i)})
			}.
			\label{eq:error_ratio_xi}
		\end{equation}
		We see that   $\xi$ is large that means   the CS symbols are much less reliable than the other information symbols, as shown in 
		Fig.~\ref{fig:xi_SNR}.
		
		\subsection{ARD-Based Branch Pruning Strategy}
		\label{construction}

		ABP-NBSCL processes the information symbols in two different ways. 
		If $i\in\mathcal{A}^r$, the symbol is regarded as reliable and the decoder makes a hard decision without path splitting. 
		If $i\in\mathcal{A}^u$, the symbol is seen as a branch-search position. 
		In this case, the decoder then evaluates the $q$ possible branches, removes branches with large accumulated reliability deviation (ARD), and forwards only the retained branches to path-metric sorting.
		
		For the $l$-th path at symbol $u_i$, let $L_l^{(i)}(\lambda)$ be the LLR metric of $\lambda\in\mathbb{G}\mathbb{F}(q)$. 
		The hard-decision symbol is decided by the locally most reliable branch, given by
		\begin{equation}
			\lambda_{l,i}^{\star}
			=
			\arg\min_{\lambda\in\mathbb{G}\mathbb{F}(q)}
			L_l^{(i)}(\lambda).
			\label{eq:local_opt_symbol}
		\end{equation}
		For a candidate branch labeled by $\lambda$, define its one-step reliability deviation as
		\begin{equation}
			\Delta_{l,i}(\lambda)
			=
			L_l^{(i)}(\lambda)-L_l^{(i)}(\lambda_{l,i}^{\star})\ge0.
			\label{eq:one_step_deviation}
		\end{equation}
		This value measures the extra metric cost of choosing $\lambda$ instead of the local hard decision. 
		The accumulated reliability deviation (ARD) predicted for this branch is
		\begin{equation}
			\tilde{D}_l^{(i)}(\lambda)
			=
			\begin{cases}
				D_l^{(i-1)}+\Delta_{l,i}(\lambda), & i\in\mathcal{A}^u,\\
				D_l^{(i-1)}, & i\notin\mathcal{A}^u.
				\label{eq:ard_predict}
			\end{cases}
		\end{equation}

		This  value $\tilde{D}_l^{(i)}(\lambda)$ is compared with the ARD threshold $\rho$. 
		If $\tilde{D}_l^{(i)}(\lambda)>\rho$, the branch is pruned. 
		Otherwise, the branch is retained and sent to PM sorting. 
		
		After a branch is retained, its ARD is updated by
		\begin{equation}
			D_l^{(i)}
			=
			\begin{cases}
				\tilde{D}_l^{(i)}(\hat{u}_{i,l}), & i\in\mathcal{A}^u,\\
				D_l^{(i-1)}, & i\in\mathcal{A}^r\cup\mathcal{A}^c.
			\end{cases}
			\label{eq:ard_update}
		\end{equation}

		The branch pruning rule is written as
		\begin{equation}
			\mathcal{Q}_{l,i}
			=
			\{\lambda\in\mathbb{G}\mathbb{F}(q):
			\lambda=\lambda_{l,i}^{\star}\ \text{or}\ 
			\tilde{D}_l^{(i)}(\lambda)\le\rho\}.
			\label{eq:valid_branch_set}
		\end{equation}
		Only symbols in $\mathcal{Q}_{l,i}$ are extended. 
		The locally best branch is always kept. 
		Branches with large accumulated deviation are removed before PM sorting.
		
		\subsection{Threshold Selection}
		
		The ARD threshold  $\rho$ is computed offline by GA. 
		For each design-SNR $\Gamma_{\mathrm{d}}$, GA provides the LLR statistics of the CS symbols. 
		The expected one-step branch deviation is then evaluated for each $i\in\mathrm{CS}$, and the largest value is used as the pruning threshold:
		\begin{equation}
			\begin{aligned}
				\bar{\Delta}_{i}(\Gamma_{\mathrm{d}})
				&=
				\mathbb{E}_{\Gamma_{\mathrm{d}}}\!\left[
				\Delta_{l,i}(\lambda)
				\mid i\in\mathrm{CS},\lambda\ne\lambda_{l,i}^{\star}
				\right],\\
				\rho(\Gamma_{\mathrm{d}})
				&=\max_{i\in\mathrm{CS}}\bar{\Delta}_{i}(\Gamma_{\mathrm{d}}).
			\end{aligned}
			\label{eq:rho_calculation}
		\end{equation}
		Thus, $\rho(\Gamma_{\mathrm{d}})$ represents the maximum average deviation that a retained non-optimal branch is allowed to accumulate at CS symbols. 
		A smaller $\Gamma_{\mathrm{d}}$ gives a tighter threshold and removes more branches, while a larger $\Gamma_{\mathrm{d}}$  keeps more branches. 
		During online decoding, the selected $\rho(\Gamma_{\mathrm{d}})$ is fixed and no additional statistical search is required. 
		%
		%
		
		The overall decoding process of ABP-NBSCL is summarized in Algorithm~\ref{alg:abp_nbscl}.
		\begin{algorithm}[ht]
			\algtext*{EndIf}
			\algtext*{EndFor}
			\caption{Adaptive Branch Pruning-Aided NB-SCL (ABP-NBSCL) Decoder}
			\label{alg:abp_nbscl}
			\begin{algorithmic}[1]
				\Require List size $L$, branch-search set $\mathcal{A}^u$, ARD threshold $\rho$, counter threshold $\omega$, and received LLRs.
				\Ensure Estimated codeword $\hat{\mathbf{x}}$.
				\State Initialize one path with $\mathrm{PM}=0$, $D_l^{(0)}=0$, and $w_l[0]=0$.
				\For{$i=1$ to $N$}
				\ForAll{surviving paths $l$}
				\If{$i\in\mathcal{A}^c$}
				\State Set $\hat{u}_{i,l}=0$, update the PM, and keep $D_l^{(i)}=D_l^{(i-1)}$.
				\ElsIf{$i\notin\mathcal{A}^u$}
				\State Compute $\lambda_{l,i}^{\star}$ by \eqref{eq:local_opt_symbol}.
				\State Set $\hat{u}_{i,l}=\lambda_{l,i}^{\star}$ without splitting, update the PM, and keep $D_l^{(i)}=D_l^{(i-1)}$.
				\State Set $w_l[i]=w_l[i-1]+1$.
				\Else
				\State Compute $\lambda_{l,i}^{\star}$ by \eqref{eq:local_opt_symbol}.
				\State Initialize $\mathcal{Q}_{l,i}=\{\lambda_{l,i}^{\star}\}$.
				\ForAll{$\lambda\in\mathbb{G}\mathbb{F}(q)\setminus\{\lambda_{l,i}^{\star}\}$}
				\State Compute $\Delta_{l,i}(\lambda)$ by \eqref{eq:one_step_deviation}.
				\State Compute $\tilde{D}_l^{(i)}(\lambda)$ by \eqref{eq:ard_predict}.
				\If{$\tilde{D}_l^{(i)}(\lambda)\le\rho$}
				\State Add $\lambda$ into $\mathcal{Q}_{l,i}$.
				\EndIf
				\EndFor
				\State Extend path $l$ only over $\lambda\in\mathcal{Q}_{l,i}$, and update the PM and ARD by \eqref{eq:ard_update}.
				\State If $|\mathcal{Q}_{l,i}|=1$, set the child counter to $w_l[i-1]+1$; otherwise reset it to zero.
				\EndIf
				\EndFor
				\If{the number of surviving paths is larger than $L$}
				\State Let $\mathcal{R}_i=\{l:w_l[i]>\omega\}$.
				\If{$\mathcal{R}_i\neq\emptyset$}
				\State Retain paths in $\mathcal{R}_i$ first and fill the remaining positions by the smallest PMs.
				\Else
				\State Retain the $L$ paths with the smallest PMs.
				\EndIf
				\EndIf
				\EndFor
				\State Output the candidate codeword with the smallest PM.
			\end{algorithmic}
		\end{algorithm}
		
		\subsection{Complexity Analysis}

		The NB-SCL decoder expands each active path into $q$ candidates at an information symbol with the complexity of $\mathcal{O}(qLN\log N)$, where $L$ is the list size and $N$ is the code length.
		Given a field order $q$, the dominant factor is the number of active paths.
		Define $L_i$ as the average number of decoding paths that are split  when $u_i$ is processed.
		In NB-SCL, each active path is split into $q$ branches, so the number of candidate paths becomes $qL_i$. 
		Before pruning, the path-growth increment is $(q-1)L_i$. 
		After pruning, at most $L$ paths survive.
		Once the path number reaches $L$, the subsequent information symbols are decoded with this size $L$.
		
		Recall that SR-NBSCL skips path splitting at reliable information symbols. 
		Let $\bar{L}_{\mathrm{SR}}$ denote the average number of active paths in SR-NBSCL.
		Given a field order $q$, its average complexity is
		
		\begin{equation}
			C_{\mathrm{SR}}
			=
			\mathcal{O}(q\bar{L}_{\mathrm{SR}}N\log N),
			\label{eq:sr_average_complexity}
		\end{equation}
		where $\bar{L}_{\mathrm{SR}}\le L$. 
		If a symbol is judged unreliable, SR-NBSCL still expands each active path into $q$ branches.
		
		ESR-NBSCL further reduces the list-decoding interval. 
		After the  index $N-K_1+1$, the final $K_1$ information symbols are decoded by SC decoding instead of list decoding. 
		Thus, the list-decoding part is shortened from $N$ symbols to approximately $N-K_1$ symbols, and the tail part is processed with SC decoding complexity. 
		Using $\bar{L}_{\mathrm{ESR}}$ to denote the average active list size, the average complexity can be expressed as
		\begin{equation}
			C_{\mathrm{ESR}}
			=
			\mathcal{O}\!\left(q\bar{L}_{\mathrm{ESR}}(N-K_1)\log N
			+q\bar{L}_{\mathrm{ESR}}K_1\log K_1\right).
			\label{eq:esr_average_complexity}
		\end{equation}
		Although ESR-NBSCL removes the tail list expansion, the remaining unreliable positions before $N-K_1+1$ may still cause rapid path growth.
		
		For $i\in\mathcal{A}^r$, the decoder makes a hard decision and no path splitting is performed.
		For $i\in\mathcal{A}^u$, only the branches in $\mathcal{Q}_{l,i}$ are retained.
		Define the effective branch factor at $u_i$ as
		\begin{equation}
			\beta_i=
			\begin{cases}
				1, & i\in\mathcal{A}^r,\\
				\frac{1}{L_i}\sum\limits_{l\in\mathcal{L}_{i-1}}|\mathcal{Q}_{l,i}|, & i\in\mathcal{A}^u,
			\end{cases}
			\qquad 1\le\beta_i\le q .
			\label{eq:effective_branch_factor}
		\end{equation}

		Thus, the path-growth increment is reduced from $(q-1)L_i$ in NB-SCL to $(\beta_i-1)L_i$ in the ABP-NBSCL.
		When $\beta_i<q$, fewer paths are generated before pruning.
		
		\begin{figure*}[!t]
			\centering
			\begin{minipage}[b]{0.325\textwidth}
				\centering
				\includegraphics[width=\linewidth]{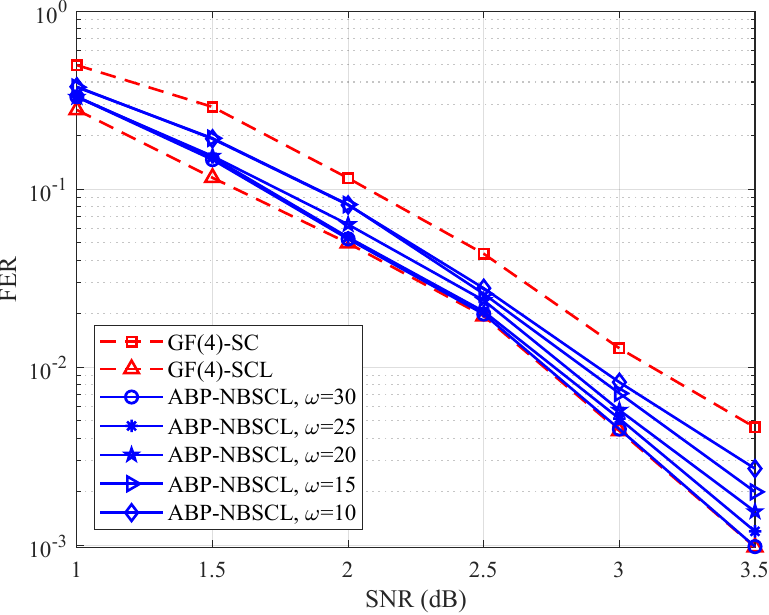}
				\vspace{0.05cm}
				\centerline{(a)}
			\end{minipage}
			\hfill
			\begin{minipage}[b]{0.325\textwidth}
				\centering
				\includegraphics[width=\linewidth]{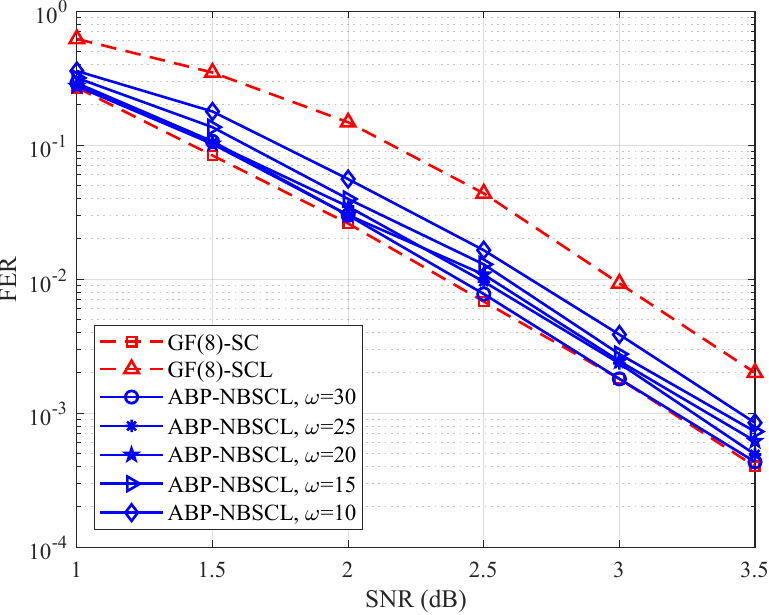}
				\vspace{0.05cm}
				\centerline{(b)}
			\end{minipage}
			\hfill
			\begin{minipage}[b]{0.325\textwidth}
				\centering
				\includegraphics[width=\linewidth]{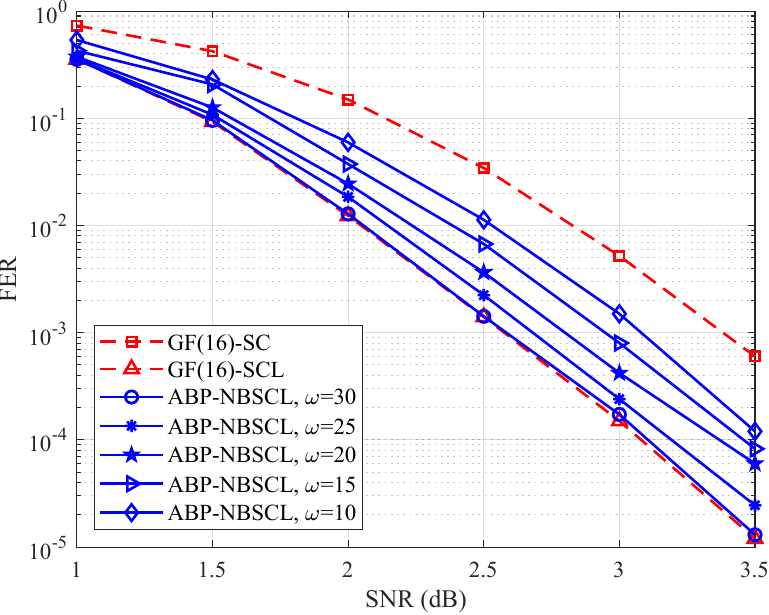}
				\vspace{0.05cm}
				\centerline{(c)}
			\end{minipage}
			\vspace{-0.15cm}
			\caption{\label{fig:fer_field_order} FER performance of the proposed ABP-NBSCL decoder with $N=128$, $R=0.5$, and $L=8$: (a) over $\mathbb{G}\mathbb{F}(4)$, (b) over $\mathbb{G}\mathbb{F}(8)$, and (c) over $\mathbb{G}\mathbb{F}(16)$.}
			\vspace{-0.45cm}
		\end{figure*}
		\begin{figure*}[!t]
			\centering
			\begin{minipage}[b]{0.325\textwidth}
				\centering
				\includegraphics[width=\linewidth]{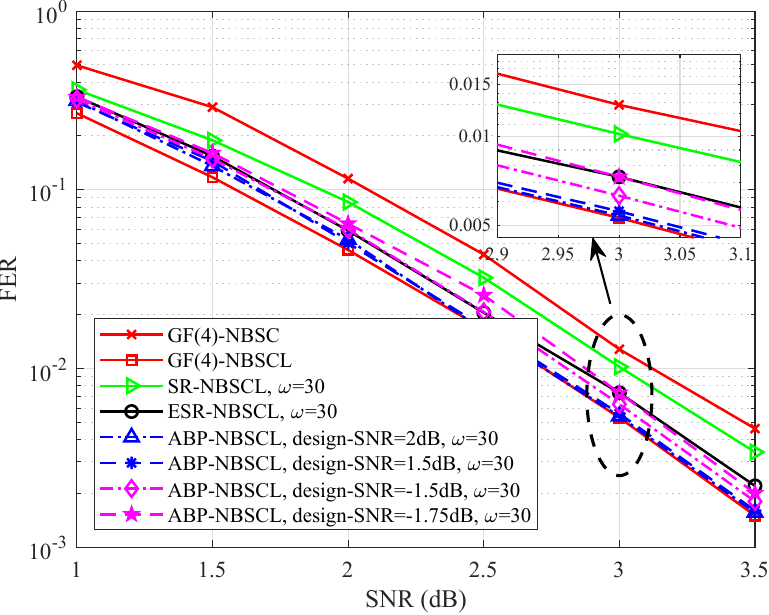}
				\vspace{0.05cm}
				\centerline{(a)}
			\end{minipage}
			\hfill
			\begin{minipage}[b]{0.325\textwidth}
				\centering
				\includegraphics[width=\linewidth]{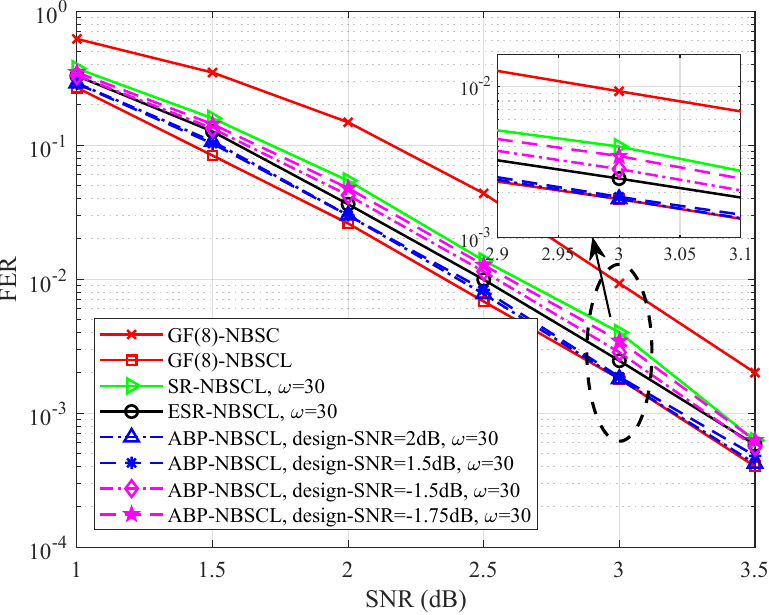}
				\vspace{0.05cm}
				\centerline{(b)}
			\end{minipage}
			\hfill
			\begin{minipage}[b]{0.325\textwidth}
				\centering
				\includegraphics[width=\linewidth]{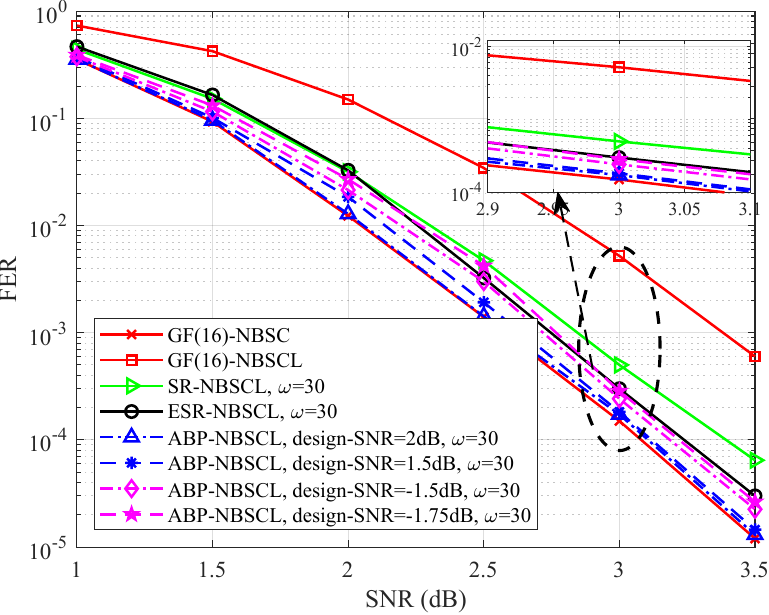}
				\vspace{0.05cm}
				\centerline{(c)}
			\end{minipage}
			\vspace{-0.15cm}
			\caption{\label{fig:fer_abp} FER performance comparison of ABP-NBSCL under different design-SNRs with $L=8$: (a) over $\mathbb{G}\mathbb{F}(4)$, (b) over $\mathbb{G}\mathbb{F}(8)$, and (c) over $\mathbb{G}\mathbb{F}(16)$.}
			\vspace{-0.5cm}
		\end{figure*}
		
		\begin{figure}[!t]
			\centering
			\includegraphics[width=0.95\columnwidth]{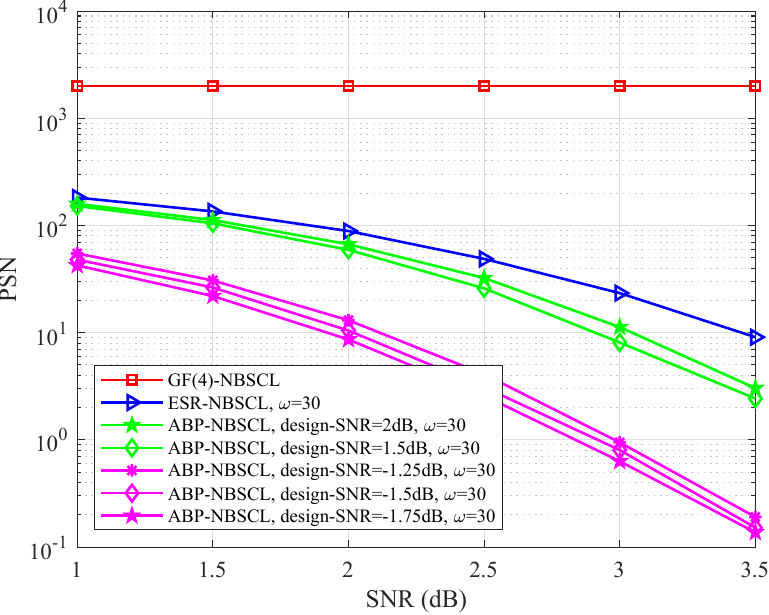}
			\vspace{-0.15cm}
			\caption{\label{fig:psn_comparison} PSN performance of different algorithms for an NBPC with $L=8$.}
			\vspace{-0.5cm}
		\end{figure}
		
		\begin{figure}[!t]
			\centering
			\includegraphics[width=0.9\columnwidth]{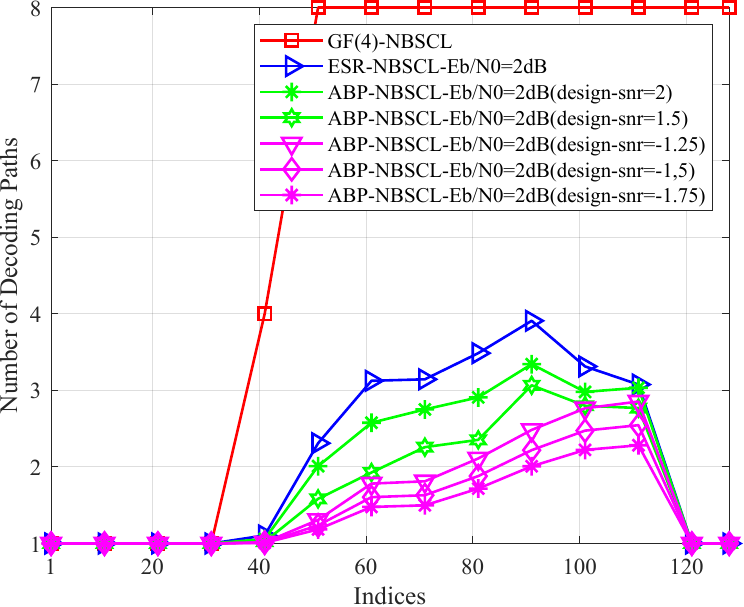}
			\vspace{-0.15cm}
			\caption{\label{fig:paths_index_compare} Comparison of the average number of surviving paths at each decoding index for different decoding algorithms.}
			\vspace{-0.5cm}
		\end{figure}
		
		\begin{figure}[!t]
			\centering
			\includegraphics[width=0.95\columnwidth]{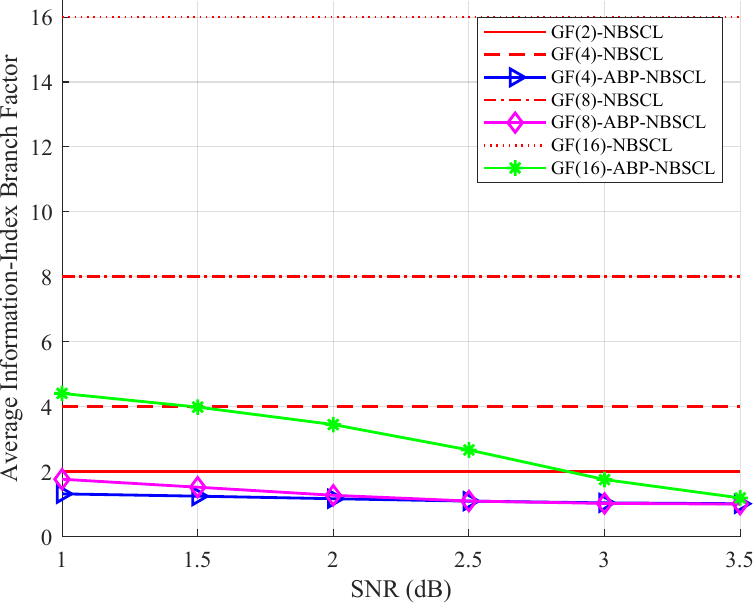}
			\vspace{-0.15cm}
			\caption{\label{fig:beta_compare} Comparison of the effective branch factor $\beta$ for NB-SCL and ABP-NBSCL decoders.}
			\vspace{-0.45cm}
		\end{figure}
		
		Let $\bar{L}_{\mathrm{ABP}}$ be the average number of active paths in ABP-NBSCL.
		Let $\bar{\beta}$ be the average effective branch factor.
		The decoding complexity is
		\begin{equation}
			C_{\mathrm{ABP}}
			=
			\mathcal{O}\!\left(\bar{\beta}\bar{L}_{\mathrm{ABP}}N\log N\right),
			\label{eq:abp_average_complexity}
		\end{equation}

		We see that as compared with \eqref{eq:sr_average_complexity}  
		for  SR-NBSCL, the list size  is replaced by the   lower  list size $\bar{\beta}\bar{L}_{\mathrm{ABP}}$, since  $\beta_i\leq q$ and $\bar{L}_{\mathrm{ABP}}< \bar{L}_{\mathrm{SR}}$  .

		\section{Simulation Results}
		\label{simulation}
		
		This section shows simulation results  of frame error rate (FER) and complexity of different NB-SCL schemes.
		Since ABP-NBSCL gives the best FER-complexity tradeoff among SR-NBSCL, ESR-NBSCL, and ABP-NBSCL, SR-NBSCL and ESR-NBSCL are used only as reference curves.
		The code length is $N=128$, the rate is $R=0.5$, and the list size is $L=8$.
		The codewords are modulated by binary phase-shift keying and transmitted over an additive white Gaussian noise ) channel.
		The NBPCs are constructed by the Monte-Carlo method.
		The primitive polynomials for $\mathbb{G}\mathbb{F}(4)$ and $\mathbb{G}\mathbb{F}(8)$ are $f(x)=x^2+x+1$ and $f(x)=x^3+x+1$, respectively.
		
		Fig.~\ref{fig:fer_field_order} shows the FER performance of SC, SCL, and ABP-NBSCL decoders over $\mathbb{G}\mathbb{F}(4)$, $\mathbb{G}\mathbb{F}(8)$, and $\mathbb{G}\mathbb{F}(16)$, where all ABP-NBSCL curves are obtained with design-SNR $=2$ dB. Recall that a  larger   $\omega$ allows more decoding trial before pruning paths. Thus, as $\omega$ increases, i.e., $\omega=30$, the performance of ABP-NBSCL improves and becomes closer to that of NB-SCL. 
		These results show that the proposed pruning rule reduces path expansion without degrading the FER performance. 
		
		Fig.~\ref{fig:fer_abp} compares ABP-NBSCL with NB-SC, NB-SCL, SR-NBSCL, and ESR-NBSCL over $\mathbb{G}\mathbb{F}(4)$, $\mathbb{G}\mathbb{F}(8)$ and $\mathbb{G}\mathbb{F}(16)$.
		We can see that the ABP-NBSCL curve with  the ARD threshold $\rho=20.96$ in \eqref{eq:rho_calculation} that is obtained for the design-SNR $2\mathrm{dB}$,   performs almost the same as  NB-SCL, especially at the high SNR. Moreover,  ABP-NBSCL also has close performance to ESR-NBSCL   in FER and outperforms   SR-NBSCL.

		To further provide a quantitative comparison of complexity, we use the total PSN and the average number of decoding paths to compare  the different NB-SCL schemes in simulations, which is positively proportional to both  the number of active path and the effective branch, as  given in \eqref{eq:sr_average_complexity}--\eqref{eq:abp_average_complexity}.  
		Let $T$ be the total number of simulation trials, The total PSN in the decoding is $P_{\mathrm{SN}}=\frac{1}{T}\sum_{j=1}^{T}\sum_{i\in\mathcal{A}}L_{i,j}$. The average number of decoding paths $L_i=\frac{1}{T}\sum_{j=1}^{T}L_{i,j}$, where $L_{i,j}$ is defined as the number of splitting paths at stage $i$ in the $j$-th experiment. 
		
		Fig.~\ref{fig:psn_comparison} compares the PSN of NB-SCL, ESR-NBSCL, and ABP-NBSCL with $L=8$.
		We see that NB-SCL expands the full list at information symbols, so its PSN is the highest among these schemes and changes little with SNR.
		ESR-NBSCL lowers the PSN when the SNR increases, but it still relies on the non-splitting counter to reduce paths.
		ABP-NBSCL has the smallest PSN at all tested SNRs because ARD pruning removes branches before list sorting.
		At $\mathrm{SNR}= 3 \mathrm{dB}$, the PSN of ESR-NBSCL is about $30$, while the ABP-NBSCL   with design-SNRs $ 2 \mathrm{dB}$ and $-1.75\mathrm{dB}$ has PSN of only 10  and $1$, respectively.
		This gives more than $80\%$ reduction in active path splitting.
		
		Fig.~\ref{fig:paths_index_compare} compares the average number of active paths $L_i$ at each decoding index for an NBPC with $L=8$ and $E_b/N_0=2\mathrm{dB}$. 
		NB-SCL reaches the full list size $L=8$ soon after path splitting starts and keeps this list size for most decoding indices. 
		ESR-NBSCL reduces the path number by using the split-reduced rule and NB-SC decoding, and its peak value is about $4$. In particular, 
		ABP-NBSCL gives a slower path growth because the branches with large ARD values are pruned before PM sorting. 
		For the tested design-SNRs, the peak average path number of ABP-NBSCL is about 3, which is about $25\%$ lower than ESR-NBSCL and more than $60\%$ lower than NB-SCL. 
		A smaller design-SNR gives a tighter ARD threshold and keeps fewer candidate branches. Thus, together with the performance comparison in Fig.~\ref{fig:fer_abp},  the ABP-NBSCL   effectively minimizes computational complexity without compromising error-rate performance. 
		
		Fig.~\ref{fig:beta_compare} plots the average effective branch factor $\bar{\beta}$ over all information-symbol indices under different SNRs.
		For NB-SCL over $\mathbb{G}\mathbb{F}(q)$, each active path is split into $q$ candidate branches when an unfrozen symbol is decoded, so its $\bar{\beta}=q$.
		For ABP-NBSCL, $\bar{\beta}$ is determined by the retained branch set $\mathcal{Q}_{l,i}$ in \eqref{eq:effective_branch_factor} and represents the average number of retained candidates per active path.
		It can be seen that the ABP-NBSCL curves stay close to $\bar{\beta}=1$ and are much smaller than $q$ at high SNR, i.e., with 3.5 dB. 
		This is because ABP-NBSCL prunes unreliable candidate branches by the ARD threshold.

		For each surviving path,  as given in \eqref{eq:local_opt_symbol}--\eqref{eq:ard_update}, ABP-NBSCL first computes the hard-decision symbol $\lambda_{l,i}^{\star}$, obtains $\tilde{D}_l^{(i)}(\lambda)$ for the other candidates, and retains only the candidates satisfying $\tilde{D}_l^{(i)}(\lambda)\le\rho$ before PM sorting.
		In Fig.~\ref{fig:beta_compare}, at a low SNR, more candidate paths are kept and $\bar{\beta}$ is larger.
		As the SNR increases, the hard-decision symbol becomes more reliable with fewer candidates, then  $\bar{\beta}$ decreases.
		In particular, its $\bar{\beta}$  is even less than $2$, which is the branch factor of binary SCL.
		According to $C_{\mathrm{ABP}}=\mathcal{O}(\bar{\beta}\bar{L}_{\mathrm{ABP}}N\log N)$, reducing $\bar{\beta}$ decreases the number of candidate branches before PM sorting and therefore lowers the decoding complexity.
		
		\section{Conclusions}
		\label{conclusion}
		In this paper, we investigate low-complexity SCL decoding for $2\times2$ kernel non-binary polar codes. To alleviate the $q$-ary branch expansion of conventional NB-SCL decoding at information symbols, we propose three low-complexity decoders, i.e., SR-NBSCL, ESR-NBSCL, and ABP-NBSCL. The SR-NBSCL decoder employs GA-based reliability thresholds to skip path splitting for highly reliable information symbols. The ESR-NBSCL decoder further exploits the final Rate-1 information-symbol block, where the SR-NBSCL splitting rule is applied before $u_{N-K_1+1}$ and simplified NB-SC decoding is used for the remaining nodes. The ABP-NBSCL decoder introduces an accumulated reliability deviation metric to prune unreliable candidate branches before path-metric sorting. Simulation results show that the proposed decoders significantly reduce the path splitting number and decoding complexity while maintaining FER performance close to that of conventional NB-SCL decoding. In particular, ABP-NBSCL achieves more than 80\% PSN reduction at several tested SNRs with negligible performance loss.

		\appendices
		
		\section{Proof of Conjecture 1}
		
		We use the all-zero codeword for the proof. This causes no loss of generality for a symmetric $q$-ary channel. Assume that the correct path reaches the $i$-th unfrozen symbol. Then $\hat{u}_1^{i-1}=u_1^{i-1}$. For a wrong symbol $\eta\in\mathbb{G}\mathbb{F}(q)\setminus\{0\}$, define the pairwise LLR as
		\begin{equation}
			\Delta_{\eta}^{(i)}
			=\log\frac{P_l(u_i=0\mid y_1^N,\hat{u}_1^{i-1})}
			{P_l(u_i=\eta\mid y_1^N,\hat{u}_1^{i-1})}.
			\label{eq:app_delta_pair}
		\end{equation}
		For the all-zero codeword, $\Delta_{\eta}^{(i)}$ compares the correct symbol $0$ with the wrong symbol $\eta$. Under the non-binary Gaussian approximation, we write
		\begin{equation}
			\Delta_{\eta}^{(i)}\sim \mathcal{N}(m_i,2m_i),
			\qquad \eta\neq0.
			\label{eq:app_ga_pair}
		\end{equation}
		For simple notation, the same mean $m_i$ is used for all pairwise LLRs. If the pairwise means are not equal, $m_i$ can be chosen as the smallest one. The same upper bounds still hold. The pairwise error probability is
		\begin{equation}
			p_i
			=\Pr\{\Delta_{\eta}^{(i)}<0\}
			=Q\left(\sqrt{\frac{m_i}{2}}\right).
			\label{eq:app_pep}
		\end{equation}
		The symbol error event is the union of the $q-1$ pairwise error events. Hence
		\begin{equation}
			P_e(u_i)
			=\Pr\left\{\bigcup_{\eta\neq0}\left\{\Delta_{\eta}^{(i)}<0\right\}\right\}
			\le (q-1)p_i .
			\label{eq:app_sep_bound}
		\end{equation}
	 Under the union-bound Gaussian approximation used in this paper, we set
		\begin{equation}
			p_i=\frac{P_e(u_i)}{q-1}=Q(t_i),
			\qquad
			t_i=Q^{-1}\left(\frac{P_e(u_i)}{q-1}\right).
			\label{eq:app_t_def}
		\end{equation}
		Then $m_i=2t_i^2$. The splitting threshold is
		\begin{equation}
			T_i=\log\frac{1-P_e(u_i)}{P_e(u_i)}
			=\log\left(\frac{1}{(q-1)Q(t_i)}-1\right).
			\label{eq:app_t_threshold}
		\end{equation}
		We first consider the event that the decoder makes a wrong decision without splitting. For a fixed wrong symbol $\eta$, this event requires
		\begin{equation}
			\Delta_{\eta}^{(i)}<-T_i .
			\label{eq:app_wrong_event}
		\end{equation}
		Thus the probability of this event is
		\begin{equation}
			P'_{i,\eta}
			=\Pr\{\Delta_{\eta}^{(i)}<-T_i\}
			=Q\left(\frac{m_i+T_i}{\sqrt{2m_i}}\right)
			=Q\left(t_i+\frac{T_i}{2t_i}\right).
			\label{eq:app_wrong_pair_prob}
		\end{equation}
		For all wrong symbols, the wrong no-splitting probability satisfies
		\begin{equation}
			P'_{w,i}\le(q-1)Q\left(t_i+\frac{T_i}{2t_i}\right).
			\label{eq:app_wrong_all_prob}
		\end{equation}
		Since $P_e(u_i)=(q-1)Q(t_i)$ under the above approximation, we have
		\begin{equation}
			\frac{P'_{w,i}}{P_e(u_i)}
			\le
			\frac{Q\left(t_i+\frac{T_i}{2t_i}\right)}
			{Q(t_i)} .
			\label{eq:app_wrong_ratio}
		\end{equation}
		When $P_e(u_i)\rightarrow0$, we have $t_i\rightarrow+\infty$. From \eqref{eq:app_t_threshold},
		\begin{equation}
			\lim_{t_i\rightarrow+\infty}T_i=+\infty,
			\qquad
			\lim_{t_i\rightarrow+\infty}\frac{T_i}{2t_i^2}=\frac{1}{4}.
			\label{eq:app_threshold_limit}
		\end{equation}
		Using the same Gaussian-tail step as the binary proof,
		\begin{equation}
			\begin{aligned}
					\frac{Q\left(t_i+\frac{T_i}{2t_i}\right)}{Q(t_i)}
					&\sim
					\frac{t_i}{t_i+\frac{T_i}{2t_i}}
					\exp\left(-\frac{T_i}{2}-\frac{T_i^2}{8t_i^2}\right).
				\end{aligned}
			\label{eq:app_tail_ratio}
		\end{equation}
		Since $T_i\rightarrow+\infty$, \eqref{eq:app_tail_ratio} gives
		\begin{equation}
			\lim_{t_i\rightarrow+\infty}
			\frac{Q\left(t_i+\frac{T_i}{2t_i}\right)}
			{Q(t_i)}
			=0.
			\label{eq:app_wrong_ratio_limit}
		\end{equation}
		Thus
		\begin{equation}
			\lim_{P_e(u_i)\rightarrow0}
			\frac{P'_{w,i}}{P_e(u_i)}=0.
			\label{eq:app_wrong_final_limit}
		\end{equation}
		Thus, the probability that a wrong symbol is selected without splitting is much smaller than the symbol error probability.
		
		Next, we consider the event that the correct path does not split. The correct path does not split when
		\begin{equation}
			\min_{\eta\neq0}\Delta_{\eta}^{(i)}>T_i.
			\label{eq:app_nosplit_correct}
		\end{equation}
		The probability that the correct path splits is
		\begin{equation}
			P_{s,i}
			=\Pr\left\{\bigcup_{\eta\neq0}\left\{\Delta_{\eta}^{(i)}\le T_i\right\}\right\}
			\le(q-1)Q\left(t_i-\frac{T_i}{2t_i}\right).
			\label{eq:app_correct_split_bound}
		\end{equation}
		From \eqref{eq:app_threshold_limit},
		\begin{equation}
			t_i-\frac{T_i}{2t_i}
			=t_i\left(1-\frac{T_i}{2t_i^2}\right)
			\rightarrow+\infty.
			\label{eq:app_correct_arg}
		\end{equation}
		Therefore
		\begin{equation}
			\lim_{P_e(u_i)\rightarrow0}P_{s,i}=0.
			\label{eq:app_split_limit}
		\end{equation}
		Let $P_{r,i}=1-P_{s,i}$ denote the probability that the correct path does not split and chooses the correct symbol. Then
		\begin{equation}
			\lim_{P_e(u_i)\rightarrow0}P_{r,i}=1.
			\label{eq:app_correct_final_limit}
		\end{equation}
		Equations \eqref{eq:app_wrong_final_limit} and \eqref{eq:app_correct_final_limit} are the $q$-ary extensions of the probability result in the binary split-reduced SCL proof. The correct path is decoded without splitting with probability close to one. A wrong no-splitting decision has a much smaller probability than $P_e(u_i)$. Since the selected symbol is the minimum-LLR symbol, the PM update in \eqref{eq:pm_update} adds no penalty to the correct path. The same argument holds at the following reliable symbols. The correct path then tends to survive until the end without splitting.
		
		\section{Proof of Conjecture 2}
		 
Consider an incorrect path that survives up to symbol $u_i$, which means that at least one previous symbol has been wrongly decoded. Suppose that the erroneous decision is passed to the left child of a node $(v,s)$. The soft information of the corresponding right child is then computed by the non-binary $G$-function. Let $\chi_k^{(2v,s-1)}$ denote the correct left-child decision and $\tilde{\chi}_k^{(2v,s-1)}$ denote the decision carried by the incorrect path. If the correct decision is used, the right-child LLR for symbol $\lambda$ is given by
		\begin{equation}
			\begin{aligned}
					l_{k+2^{s-1},\lambda}^{(2v+1,s-1)}
					\approx&\; l_{k,\chi_k^{(2v,s-1)}+\gamma\lambda}^{(v,s)}
					+l_{k+2^{s-1},\delta\lambda}^{(v,s)}\\
					&-l_{k,\chi_k^{(2v,s-1)}}^{(v,s)}
					-l_{k+2^{s-1},0}^{(v,s)} .
				\end{aligned}
			\label{eq:app_g_correct}
		\end{equation}
		Let the erroneous left-child decision be $\chi_k^{(2v,s-1)}+e$, where $e\in\mathrm{GF}(q)\setminus\{0\}$. Then, the incorrect path computes the right-child LLR for symbol $\lambda$ as 
		\begin{equation}
			\begin{aligned}
					\tilde{l}_{k+2^{s-1},\lambda}^{(2v+1,s-1)}
					\approx&\; l_{k,\chi_k^{(2v,s-1)}+e+\gamma\lambda}^{(v,s)}
					+l_{k+2^{s-1},\delta\lambda}^{(v,s)}\\
					&-l_{k,\chi_k^{(2v,s-1)}+e}^{(v,s)}
					-l_{k+2^{s-1},0}^{(v,s)} .
				\end{aligned}
			\label{eq:app_g_wrong}
		\end{equation}

		Compared with \eqref{eq:app_g_correct}, \eqref{eq:app_g_wrong} contains a nonzero shift $e$ in the first and third terms. Since $\gamma\neq0$, $\lambda\mapsto e+\gamma\lambda$ is a bijection over $\mathrm{GF}(q)$. Thus, the LLR vector is permuted by a nonzero field offset and becomes mismatched with the correct conditional channel.

		The incorrect previous decision is further propagated to the descendant nodes. If a child node contains $m$ erroneous symbols, at least $m$ descendant leaf LLR vectors become mismatched. This can be shown by induction on the decoding tree, since the $F$-function combines all $q$ hypotheses, while the $G$-function preserves a one-to-one field shift due to $\gamma,\delta\neq0$.
		
		Now consider one mismatched descendant symbol $u_j$, where $j>i$. For the incorrect path, define the pairwise LLR between the symbol selected by this path, $\hat{u}_j$, and a competing symbol $\eta$ as
		\begin{equation}
			\widetilde{\Delta}_{\eta}^{(j)}
			=
			\log\frac{P_l(u_j=\hat{u}_j\mid y_1^N,\hat{u}_1^{j-1})}
			{P_l(u_j=\eta\mid y_1^N,\hat{u}_1^{j-1})},
			\qquad \eta\neq\hat{u}_j .
			\label{eq:app_wrong_delta}
		\end{equation}
		Under the Gaussian approximation, the mismatch caused by \eqref{eq:app_g_wrong} reduces the mean of at least one pairwise LLR. Denote one such pairwise LLR by $\widetilde{\Delta}_{\eta_0}^{(j)}$. We write
		\begin{equation}
			\widetilde{\Delta}_{\eta_0}^{(j)}
			\sim\mathcal{N}(\widetilde{m}_j,2\widetilde{m}_j),
			\qquad
			\widetilde{m}_j=o(m_j),
			\label{eq:app_wrong_ga}
		\end{equation}
		where $m_j$ is the corresponding mean on the correct path. Let
		\begin{equation}
			\widetilde{t}_j=\sqrt{\frac{\widetilde{m}_j}{2}},
			\qquad
			t_j=\sqrt{\frac{m_j}{2}}.
			\label{eq:app_wrong_t}
		\end{equation}
		Then $\widetilde{t}_j/t_j\rightarrow0$ when the correct sub-channel becomes reliable.
		
		Use the same $q$-ary splitting threshold at symbol $u_j$,
		\begin{equation}
			T_j
			=\log\frac{1-P_e(u_j)}{P_e(u_j)}
			=\log\left(\frac{1}{(q-1)Q(t_j)}-1\right).
			\label{eq:app_wrong_threshold_def}
		\end{equation}
		The incorrect path does not split at $u_j$ only if all pairwise LLRs exceed the threshold $T_j$.
		\begin{equation}
			\widetilde{P}_{ns,j}
			=\Pr\left\{\min_{\eta\neq\hat{u}_j}
			\widetilde{\Delta}_{\eta}^{(j)}>T_j\right\}
			\le
			\Pr\left\{\widetilde{\Delta}_{\eta_0}^{(j)}>T_j\right\}.
			\label{eq:app_wrong_nosplit_event}
		\end{equation}
		Using \eqref{eq:app_wrong_ga}, we have
		\begin{equation}
			\widetilde{P}_{ns,j}
			\le
			Q\left(\frac{T_j-\widetilde{m}_j}{\sqrt{2\widetilde{m}_j}}\right)
			=
			Q\left(\frac{T_j}{2\widetilde{t}_j}-\widetilde{t}_j\right).
			\label{eq:app_wrong_nosplit_prob}
		\end{equation}
		From Appendix~A, the correct-path threshold satisfies
		\begin{equation}
			\lim_{t_j\rightarrow+\infty}\frac{T_j}{2t_j^2}=\frac{1}{4}.
			\label{eq:app_wrong_threshold}
		\end{equation}
		Since $\widetilde{t}_j/t_j\rightarrow0$,
		\begin{equation}
			\frac{T_j}{2\widetilde{t}_j}-\widetilde{t}_j
			=
			t_j\left(
			\frac{T_j}{2t_j^2}\frac{t_j}{\widetilde{t}_j}
			-\frac{\widetilde{t}_j}{t_j}
			\right)
			\rightarrow+\infty.
			\label{eq:app_wrong_arg_limit}
		\end{equation}
		Thus
		\begin{equation}
			\lim_{P_e(u_j)\rightarrow0}\widetilde{P}_{ns,j}=0.
			\label{eq:app_wrong_nosplit_limit}
		\end{equation}
		Let $\widetilde{P}_{s,j}=1-\widetilde{P}_{ns,j}$ be the probability that the incorrect path splits at symbol $u_j$. Then
		\begin{equation}
			\lim_{P_e(u_j)\rightarrow0}\widetilde{P}_{s,j}=1.
			\label{eq:app_wrong_split_limit}
		\end{equation}

		The wrong previous decision causes error propagation to descendant symbols.
		Hence, by \eqref{eq:app_wrong_nosplit_limit}, the incorrect path splits at a later decoding stage with probability tending to one.
		
		\section{Proof of Theorem 1}
		
		We first give the ML rule for a $q$-ary code. For a codeword $\mathbf{x}=(x_1,\ldots,x_N)\in\mathcal{C}$, define
		\begin{equation}
			L_t(\lambda)=\log\frac{P(y_t\mid0)}{P(y_t\mid\lambda)},
			\qquad \lambda\in\mathbb{G}\mathbb{F}(q).
			\label{eq:app_qary_llr}
		\end{equation}
		The ML decoder gives
		\begin{align}
			\hat{\mathbf{x}}
			&=\arg\max_{\mathbf{x}\in\mathcal{C}} P(y_1^N\mid\mathbf{x}) \notag\\
			&=\arg\max_{\mathbf{x}\in\mathcal{C}}\sum_{t=1}^{N}\log P(y_t\mid x_t) \notag\\
			&=\arg\min_{\mathbf{x}\in\mathcal{C}}\sum_{t=1}^{N}L_t(x_t).
			\label{eq:app_qary_ml}
		\end{align}
		The last step holds because $\sum_{t=1}^{N}\log P(y_t\mid0)$ is fixed.
		
		Let the last $K_1=2^{k_1}$ consecutive unfrozen symbols form the Rate-1 node $B$. Assume that all previous unfrozen symbols $u_1^{N-K_1}$ are given correctly. Then the SC update gives a fixed LLR matrix at node $B$,
		\begin{equation}
			\mathbf{L}_B=[L_{B,0},L_{B,1},\ldots,L_{B,K_1-1}].
			\label{eq:app_lb}
		\end{equation}
		Let
		\begin{equation}
			\mathbf{s}=(u_{N-K_1+1},\ldots,u_N)
			\in\mathbb{G}\mathbb{F}(q)^{K_1}
			\label{eq:app_suffix}
		\end{equation}
		denote an estimate of the remaining $K_1$ unfrozen symbols.
		The corresponding codeword of node $B$ is
		\begin{equation}
			\mathbf{b}=\mathbf{s}^{T}G_2^{\otimes k_1}.
			\label{eq:app_local_transform}
		\end{equation}
		Because $\det(G_2)=\mu\delta\neq0$, the matrix $G_2^{\otimes k_1}$ is nonsingular over $\mathbb{G}\mathbb{F}(q)$. Thus $\mathbf{s}\mapsto\mathbf{b}$ is one-to-one.
		
		By \eqref{eq:app_qary_ml}, the ML decoder selects the estimate of $(u_{N-K_1+1},\ldots,u_N)$ as
		\begin{equation}
			\hat{\mathbf{s}}_{\mathrm{ML}}
			=
			\arg\min_{\mathbf{s}\in\mathbb{G}\mathbb{F}(q)^{K_1}}
			\sum_{t=0}^{K_1-1}L_{B,t}(b_t).
			\label{eq:app_ml_suffix_new}
		\end{equation}
		Since node $B$ is a Rate-1 node, $\mathbf{b}$ can be any vector in $\mathbb{G}\mathbb{F}(q)^{K_1}$. Hence
		\begin{equation}
			\hat{\mathbf{b}}
			=
			\arg\min_{\mathbf{b}\in\mathbb{G}\mathbb{F}(q)^{K_1}}
			\sum_{t=0}^{K_1-1}L_{B,t}(b_t).
			\label{eq:app_ml_b}
		\end{equation}
		The metric in \eqref{eq:app_ml_b} is a sum of independent symbol metrics. So each symbol is chosen by
		\begin{equation}
			\hat{b}_t=\arg\min_{\lambda\in\mathbb{G}\mathbb{F}(q)}L_{B,t}(\lambda),
			\qquad 0\le t\le K_1-1.
			\label{eq:app_b_hd}
		\end{equation}
		The simplified NB-SC decoder for a Rate-1 node does exactly \eqref{eq:app_b_hd}. It then applies $(G_2^{\otimes k_1})^{-1}$ to get $\hat{\mathbf{s}}$. 
		
		\section{Proof of Theorem 2}

Before reaching $u_{N-K_1+1}$, ESR-NBSCL performs the same operations as SR-NBSCL. 
Suppose that $\ell$ paths survive at this switching index. 
For each surviving prefix path, the remaining symbols $(u_{N-K_1+1},\ldots,u_N)$ form a terminal Rate-1 non-binary subcode with $q^{K_1}$ possible suffix estimates. 
If SR-NBSCL continues list decoding, these suffixes are generated through symbol-by-symbol splitting and pruning. 
In contrast, according to Theorem~1, simplified NB-SC decoding directly obtains the best suffix estimate for each surviving prefix path.

Therefore, ESR-NBSCL keeps the same prefix paths as SR-NBSCL before $u_{N-K_1+1}$, but replaces the suffix list expansion with the best NB-SC suffix decision. 
After all surviving paths are processed, the candidate codeword with the smallest path metric is selected. 
Hence, the decoding error performance of ESR-NBSCL is no worse than that of SR-NBSCL.
		
		\section{Proof of Theorem 3}
		
		Let $\mathbf{s}=(s_0,s_1,\ldots,s_{M-1})\in\mathbb{G}\mathbb{F}(q)^M$ be the local information vector, and let
		\begin{equation}
			\mathbf{b}=\mathbf{s}^{T}G_2^{\otimes m}
		\end{equation}
		be the local codeword. 
		Let the hard-decision error vector be $\mathbf{e}=(e_0,e_1,\ldots,e_{M-1})$, so that $\hat{\mathbf{b}}=\mathbf{b}+\mathbf{e}$. 
		Since $\det(G_2)=\mu\delta\neq0$, $G_2^{\otimes m}$ is nonsingular over $\mathbb{G}\mathbb{F}(q)$. 
		Hence
		\begin{equation}
			\hat{\mathbf{s}}-\mathbf{s}
			=
			\mathbf{e}^{T}\left(G_2^{\otimes m}\right)^{-1}.
		\end{equation}
		Let $a_t=\left[\left(G_2^{\otimes m}\right)^{-1}\right]_{t,0}$. 
		The first column of $G_2^{-1}$ has no zero entry because $\mu,\gamma,\delta\neq0$. 
		The Kronecker product preserves this property, so $a_t\neq0$ for all $0\le t\le M-1$. 
		The first-symbol error event is therefore
		\begin{equation}
			\mathcal{E}_{0}
			=
			\left\{
			\sum_{t=0}^{M-1}a_t e_t\neq0
			\right\}.
		\end{equation}
		If $\mathrm{wt}(\mathbf{e})=1$, then the above event must hold. 
		Thus, a node error without a first-symbol error needs at least two local codeword-symbol errors:
		\begin{equation}
			\mathcal{E}_{\mathrm{R1}}\setminus\mathcal{E}_{0}
			\subseteq
			\{\mathrm{wt}(\mathbf{e})\ge2\}.
		\end{equation}
		Since the local errors are independent with probability $p$,
		\begin{equation}
			P\{\mathrm{wt}(\mathbf{e})\ge2\}
			=
			\sum_{r=2}^{M}
			\binom{M}{r}p^r(1-p)^{M-r}.
		\end{equation}
		Then, we can have  the bound in Theorem~3. 
		As $\epsilon\rightarrow0$, the upper bound approaches zero. 
		This completes the proof.
		
		\bibliographystyle{IEEEtran}
		\bibliography{refNew}
		
	\end{document}